\documentclass[twocolumn,showpacs,preprintnumbers,amsmath,amssymb,amsfonts, final, a4paper,aps, citeautoscript,footinbib,prb,superscriptaddress]{revtex4-1}
\usepackage[pdftex]{graphicx,color}
\usepackage[latin1]{inputenc}
\usepackage{mathrsfs}
\usepackage{epstopdf}
\usepackage{times}
\usepackage{float} 
\usepackage[vcentermath]{youngtab}

\usepackage[colorlinks,bookmarks=true,citecolor=blue,linkcolor=blue,urlcolor=blue, breaklinks=true]{hyperref}
\graphicspath{{./Figs/}}

\usepackage{ulem}

\begin{document}
\title{Loop-current orderings in SU($N$)  two-leg fermionic ladder through duality symmetries}
\author{M. Habchy, F. Rose, and P. Lecheminant} 
\affiliation{Laboratoire de Physique Th\'eorique et
 Mod\'elisation, CNRS UMR 8089, CY Cergy Paris Université,
95302 Cergy-Pontoise Cedex, France.}
\date{\today}

\begin{abstract}
We investigate the formation of loop-current ordered phases in half-filled SU($N$) two-leg fermionic ladders.
Using a low-energy approach, we uncover the existence of non-perturbative duality symmetries relating four competing orders. Two of these orders correspond to loop-current ordered phases that spontaneously break the time-reversal symmetry and describe charge currents circulating in a staggered pattern either around the plaquettes or along the diagonals of the ladder. These unconventional phases are shown to be dual to conventional (charge and bond) density-wave phases through an exact density-current duality symmetry existing on the lattice. From a perturbative renormalization group approach, we find that these phases for $N>2$ are stabilized in a half-filled SU($N$)  two-leg Hubbard ladder with an additional SU($N$) Hund's interaction. The effect of a small doping on these phases is also discussed.
\end{abstract}
\maketitle
\section{Introduction}

Strongly correlated systems have attracted considerable attention over the years for their potential to
stabilize novel quantum phases of matter. A particularly interesting class of such exotic phases consists of states that spontaneously break  the time-reversal (${\rm T}$) symmetry through the circulation of  local charge currents in their ground states. Over the years, these phases have been rediscovered several times and are commonly referred to as orbital antiferromagnetic (OAF), staggered-flux or d-density wave phases.
 
\begin{figure}[t]
\begin{center}
\includegraphics[scale=1.2]{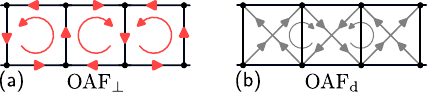}
\caption{(color online) Two possible loop-current ordered phases  which spontaneously break the time-reversal symmetry in a two-leg ladder:
(a) OAF$_{\perp}$ phase with circulating currents along the plaquettes of the ladder
(b) OAF$_{\rm d}$ phase with circulating currents along the diagonals of the ladder.}
\label{figOAFphases}
\end{center}
\end{figure}

They were first introduced by Halperin and Rice  in Ref. \onlinecite{Halperin-R-68} within the general classification of the electron-hole pairing states, and later rediscovered after the discovery of high-T$_{c}$ cuprates \cite{Affleck-M-88,Marston-A-89, Nersesyan-V-89,Schulz-89,Dombre-K-89,Hsu-M-A-91,Nayak-00,Ivanov-L-W-00,Chakravarty-al-01}. The circulating or loop-current phase breaks the translational symmetry of the underlying lattice as well as the parity symmetry. The currents around even and odd plaquettes produce local orbital moments aligned in an antiferromagnetic (staggered) way (see Fig. \ref{figOAFphases}). A different loop-current ordered  phase with broken time and parity symmetries but unbroken translational symmetry was also introduced for explaining the pseudogap phase  of high-T$_{c}$ superconductors
\cite{Varma-99}.  Polarized neutron diffraction experiments have provided compelling evidence for the existence of such a loop-current ordered phase in two-leg ladder  Sr$_{14-x}$ Ca$_x$ Cu$_{24}$ O$_{41}$ cuprate material \cite{Bounoua-al-20,Bourges-B-S-21} and in Sr$_2$IrO$_4$ compound \cite{Jeong-al-17,Murayama-al-21}.  Current ordered phases have also been predicted in Mott 
insulators with noncoplanar spin ordering \cite{Bulaevskii-al-08}.

An important question is the identification of microscopic models whose 
ground states harbor spontaneous currents.  In this respect, in a pioneering work, Nersesyan identified the simplest  model, namely a spinless two-leg ladder, which exhibits a loop-current ordered phase \cite{Nersesyan-91,Nersesyan-L-K-93}. Using the bosonization approach, he found two possible spontaneous ${\rm T}$-breaking phases in which the currents circulate either around each plaquette of the ladder or along its diagonals in a staggered pattern (see Fig. \ref{figOAFphases}). Furthermore, he gave a simple explanation for the emergence of this exotic phase
by mapping the model onto a single spinful chain \cite{Nersesyan-91,Pujari-H-09}.  After this work, loop-current ordered phases have also been predicted within the bosonization approach in various generalized spinful two-leg electronic ladders, exhibiting long-range order at half-filling and algebraically decaying correlations away from half-filling \cite{Schulz-96,Orignac-G-97,Marston-F-S-02,Fjaerestad-M-02,Tsuchiizu-F-02,Wu-L-F-03,Narozhny-C-N-05,Fjarestad-M-S-06,Chudzinski-G-G-08,Chudzinski-G-G-10,Carr-N-N-13,Beradze-N-23,Garuchava-J-N-24}. Numerical investigations of these ladders by means of the density-matrix renormalization group (DMRG) approach have reported the existence of an long-range loop-current ordered phase \cite{Marston-F-S-02,Schollwock-al-03,Fjarestad-M-S-06,Roux-O-W-P-07,Nishimoto-J-S-09}.  Such an exotic phase has also been found in higher-dimensional lattice models \cite{Capponi-W-Z-04,Kolezhuk-07,Weber-al-09}.

In this paper, we investigate the possible emergence of  loop-current ordered phases in one-dimensional (1D) multicomponent electronic systems possessing several internal degrees of freedom beyond the orbital ones. More specifically,  we consider a generalized two-leg fermionic ladder obtained by extending the usual SU(2) spin symmetry
to SU($N$). Such a continuous symmetry naturally emerges in ultracold fermionic alkaline-earth or ytterbium atoms, where the almost perfect decoupling of the electronic spin from the nuclear one in atomic collisions leads to an SU($N$) symmetry with $N=2I+1 \le 10$ ($I$ being the nuclear spin) nuclear states \cite{Cazalilla-H-U-09,Gorshkov-et-al-10,Scazza-et-al-14,Zhang-et-al-Sr-14,Cazalilla-R-14,Capponi-L-T-16,Ibarra-C-25}. 

\begin{figure}[t]
\begin{center}
\includegraphics[scale=1.1]{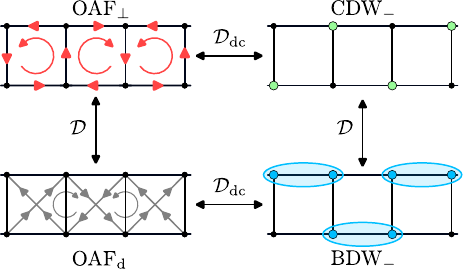}
\caption{(color online) The four competing orders related by duality symmetries;
${\cal D}_{\rm dc}$: density-current duality symmetry which is exact on the lattice; 
${\cal D}$: emergent duality symmetry in the low-energy limit.}
\label{figcompetingorderslatt}
\end{center}
\end{figure}

We derive a  low-energy approach of weakly-coupled generalized SU($N$) two-leg fermionic ladder 
at half-filling and for a generic filling close to half-filling. In particular, we find the emergence of 
four competing orders, which include the two loop-current ordered phases of Fig. \ref{figOAFphases}, and two 
conventional orders: a relative charge-density wave (CDW$_{-}$) and a longitudinal relative bond-density wave
(BDW$_{-}$) (see Fig. \ref{figcompetingorderslatt}). Interestingly, these competing orders are unified by an
emergent SO($4N$) symmetry and linked by the existence of two independent
duality symmetries: one emergent ${\cal D}$ which exists only in the low-energy limit and an exact duality
${\cal D}_{\rm dc}$ defined at the lattice level. The latter, which we refer to as the density-current duality, relates unconventional spontaneous T-breaking current phases to conventional density-wave orders as depicted 
in Fig. \ref{figcompetingorderslatt}. This duality symmetry was previously introduced in the specific $N=2$ case in
Refs. \onlinecite{Momoi-H-03,Momoi-H-05} and naturally extends to arbitrary $N$. Using a one-loop renormalization 
group (RG) approach, we show for $N>2$  that all the phases of Fig. \ref{figcompetingorderslatt} appear in the weak-coupling  regime of the half-filled SU($N$) two-leg Hubbard ladder by adding a $J_{\perp}$ interchain SU($N$) spin-spin exchange interaction, i.e., a Hund's coupling. The lightly-doped case close to half-filling is also considered using a similar approach. The loop-current ordered phases become the leading instabilities with power-law decaying correlation functions and coexist with a subleading $4k_{\rm F}$ CDW instability.  

The rest of the paper is organized as follows. In Sec. II, we introduce the lattice models considered in this work, and discuss various lattice order parameters together with the exact density-current duality ${\cal D}_{\rm dc}$  symmetry. The low-energy approach  in the weak-coupling regime at half-filling is developed in Sec. III with a special emphasis laid on 
the existence of hidden duality symmetries. The latter relate the four competing orders of 
Fig. \ref{figcompetingorderslatt} between themselves. It is also shown by a one-loop RG calculation that these four ordered phases occur for $N>2$ in the phase diagram of the half-filled SU($N$) Hubbard two-leg ladder with a Hund's interaction.
In Sec. IV, we analyse the effect of doping on the long-range ordered phases 
identified at half-filling using a one-loop RG calculation combined with an Abelian bosonization approach.
Finally, our concluding remarks are presented in Sec. V and the paper is supplied with two technical appendices.

\section{Density-current duality symmetry}
\label{sec:densitycurrduality}

In this section, we define the different lattice models that will be investigated  in this paper as well as various order parameters corresponding to possible phases. Furthermore, we  introduce the density-current duality symmetry which will play a crucial role in the following.

\subsection{The lattice models and symmetries}
\label{sec:model}

We consider a generalized SU($N$) two-leg fermionic ladder with lattice Hamiltonian: 
\begin{equation}
 {\rm H}  = {\rm H}_0  + {\rm H}_{\mathrm{int}} .
\label{eqn:hamlatticeladder}
\end{equation}
The non-interacting Hamiltonian $ {\rm H}_0$ describes the hopping of the fermions along the chains with coupling $t$ 
and in the transverse direction with hopping term $ t_\perp$:
\begin{eqnarray}
{\rm H}_0 &=& -t\sum_{i, l\alpha} \left( c_{l \alpha,i+1}^\dagger c_{l \alpha,i}+ {\rm H.c.}\right) \nonumber \\
&-& t_\perp \sum_{i ,\alpha} \left( c_{1\alpha,i}^\dagger c_{2 \alpha,i} +  {\rm H.c.} \right)-\mu \sum_{i} n_i,
\label{freehamlatt}
\end{eqnarray}
where $c^{\dagger}_{l\alpha,\,i}$  denotes a creation fermionic operator with  nuclear spin components $\alpha=1,\cdots,N$  on the $i$th site and $l=1,2$ leg. The occupation number or density at site $i$ is denoted by $n_i =  \sum_{l\alpha} c_{l \alpha,\,i}^\dag c_{l \alpha,\,i}$, and $\mu$ is the chemical potential.  

\begin{figure}[hbt]
\begin{center}
\includegraphics[scale=1.1]{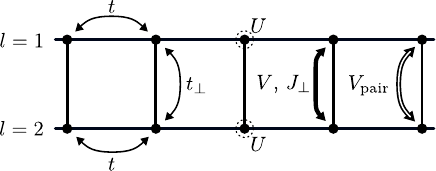}
\caption{(color online) Generalized two-leg SU($N$) ladder with longitudinal $t$ and transverse
$t_{\perp}$  hoppings, an on-site Hubbard $U$ term, an interchain density-density interaction $V$ , an
 SU($N$) Hund coupling $J_{\perp}$, and a pair-hopping term $V_{\rm pair}$.}
\label{fig2legladder}
\end{center}
\end{figure}

The interacting part of the model contains an on-site Hubbard $U$ term, an interchain density-density interaction $V$ , an
 SU($N$) Hund's coupling $J_{\perp}$, and a pair-hopping term $V_{\rm pair}$ (see Fig. \ref{fig2legladder}):
 \begin{equation}
 \begin{split}
& {\rm H}_{\mathrm{int}} = \frac{U}{2}\sum_{i}\sum_{l=1}^{2}\sum_{\alpha \neq \beta}n_{l  \alpha, i}n_{l  \beta,i}
  \\
& +V \sum_{i}  n_{1,i} n_{2,i} + J_{\perp}\sum_i   \sum_{A=1}^{N^2 -1} S^{A}_{1,i} S^{A}_{2,i}  \\
& + V_{\rm pair} \sum_i  \left( c_{1 \alpha,i}^{\dagger} 
c_{1 \beta,i}^{\dagger} c_{2 \beta,i} c_{2 \alpha,i}   + {\rm H.c.}\right),
\end{split}
 \label{doublewellint}
\end{equation}
where $n_{l \alpha, i} = c_{l \alpha,i}^\dagger c_{l \alpha,i}$, and $n_{l,i} =  \sum_{\alpha = 1}^{N} c_{l \alpha,i}^\dagger c_{l \alpha,i}$. In  Eq.  (\ref{doublewellint}), $S^{A}_{l,i}$ ($A= 1, \ldots, N^2-1$) stands for the lattice SU($N$) spin density operator of the $l=1,2$ leg:
\begin{equation}
S^{A}_{l,i} = c_{l \alpha,i}^\dagger  T^{A}_{ \alpha \beta} c_{l \beta,i}, 
 \label{SUNspindensity}
\end{equation}
where a summation over repeated greek indexes (nuclear spin degrees of freedom) is implied in the following and 
$T^{A}_{ \alpha \beta}$ are the generators in the fundamental representation of the Lie algebra of the  SU($N$) group with normalization: ${\rm Tr} ( T^{A} T^{B}) = \delta^{AB}/2$.
The interacting part (\ref{doublewellint}) can also be rewritten as:
\begin{equation}
 \begin{split}
& {\rm H}_{\mathrm{int}} = \frac{U}{2}\sum_{i}\sum_{l=1}^{2}\sum_{\alpha \neq \beta}n_{l  \alpha, i}n_{l  \beta,i}
  +\left(V - \frac{J_{\perp}}{2N} \right) \sum_{i}  n_{1,i} n_{2,i} \\
 &  - \frac{J_{\perp}}{2} \sum_i  c_{1 \alpha,i}^{\dagger} c_{2 \beta,i}^{\dagger}  c_{1 \beta,i}  c_{2 \alpha,i} 
 + V_{\rm pair} \sum_i  \left( c_{1 \alpha,i}^{\dagger} 
c_{1 \beta,i}^{\dagger}  c_{2 \beta,i}  c_{2 \alpha,i}  \right. \\
& \left. + {\rm H.c.} \right) .
\end{split}
 \label{generalint}
\end{equation}

Several known lattice models are included in the general Hamiltonian (\ref{eqn:hamlatticeladder}). When $V, J_{\perp}, V_{\rm pair} =0$ the model corresponds to the SU($N$)  generalization of the famous Hubbard two-leg ladder:
\begin{equation}
 \begin{split}
& {\rm H}_{\mathrm{hubbard}} = -t\sum_{i, l\alpha} \left( c_{l \alpha,i+1}^\dagger c_{l \alpha,i}+ {\rm H.c.}\right) 
-\mu \sum_{i} n_i  \\
& - t_\perp \sum_{i ,\alpha} \left( c_{1\alpha,i}^\dagger c_{2 \alpha,i}  
 + {\rm H.c.} \right)  + \frac{U}{2}\sum_{i}\sum_{l=1}^{2}\sum_{\alpha \neq \beta}n_{l  \alpha, i}n_{l  \beta,i},
\end{split}
 \label{hubbard2leg}
\end{equation}
where two SU($N$) Hubbard chains are coupled by a transverse hopping term. For $N=2$, this model has been extensively studied over the years as a simple description of the cuprates physics\cite{Giamarchi,Gogolin-N-T-book}.

Model (\ref{eqn:hamlatticeladder}) with $ t_\perp=0$ is also directly related to  alkaline-earth or ytterbium cold fermions 
loaded into an 1D optical lattice where the atoms can occupy the ground state ($g$ state) and the first metastable excited ($e$) state. These  two atomic states $l=g,e$ correspond to the two legs of the ladder  (\ref{eqn:hamlatticeladder}) and $\alpha, \beta= 1, \ldots, N$  label 
 the nuclear-spin multiplet.  The resulting Hamiltonian,  called the $ge$ model, which describes the low-energy collisions between two such atoms, is defined by \cite{Gorshkov-et-al-10,Cazalilla-R-14,Capponi-L-T-16,Bilokonal-23,Ibarra-C-25}
\begin{equation}
\begin{split}
& {\rm H}_{ge} = - t \sum_{i,l,\alpha} \left( c^{\dagger}_{l\alpha,\,i+1}c_{l\alpha,\,i} + {\rm H.c.}  \right) 
-\mu \sum_{i} n_i \\
& +   \frac{U_{ge}}{2}\sum_{i}\sum_{l=1}^{2}\sum_{\alpha \neq \beta}n_{l  \alpha, i}n_{l  \beta,i}
+ V_{ge} \sum_in_{g,i}n_{e,i}
 \\
&   + V_{\rm ex}\sum_{i, \alpha, \beta}c^\dagger_{g \alpha,i}c^\dagger_{e \beta,i}
c_{ g \beta,i} c_{e \alpha,i} .
\end{split}
\label{gemodel}
\end{equation}
Such lattice Hamiltonian is mapped onto model (\ref{generalint}) with $U_{ge}= U, V_{\rm ex} = - J_{\perp}/2$ and $V_{ge} = V - J_{\perp}/2N$.

A second realization of the two-leg SU($N$) fermionic ladder (\ref{eqn:hamlatticeladder}) in the context of 
cold atoms is obtained by preparing the alkaline-earth or ytterbium atoms in their $g$ state and considering a double-well optical lattice. The resulting Hamiltonian describing the low-energy properties of this double-well system is given by Eq. (\ref{generalint}) with a fine-tuning of interactions \cite{Fromholz-C-L-P-T-19}:
 \begin{equation}
 \begin{split}
& {\rm H}_{\mathrm{dw}} =   {\rm H}_0  + \frac{U}{2}\sum_{i}\sum_{l=1}^{2}\sum_{\alpha \neq \beta}n_{l  \alpha, i}n_{l  \beta,i} + V \sum_{i}  n_{1,i} n_{2,i} \\
 &  + V  \sum_i  c_{1 \alpha,i}^{\dagger} c_{2 \beta,i}^{\dagger}  c_{1 \beta,i}  c_{2 \alpha,i}  \\
 & +   \frac{V}{2}   \sum_i  \left( c_{1 \alpha,i}^{\dagger } 
c_{1 \beta,i}^{\dagger}  c_{2 \beta,i}  c_{2 \alpha,i}  + {\rm H.c.} \right) .
\end{split}
 \label{doubleintfinetuned}
\end{equation}

The general model (\ref{eqn:hamlatticeladder}) displays an U($N$) = U(1)$_{\text{c}}$ $\times$ SU($N$) continuous invariance,  where the U(1)$_{\text{c}}$ symmetry ($c_{l\alpha,i} \rightarrow e^{i \theta} c_{l\alpha,i}$) describes the conservation of the total number of fermions and the SU($N$) symmetry  ($c_{l\alpha,i} \rightarrow U_{\alpha \beta} c_{l\beta,i}$, $U$ being an SU($N$) matrix) accounts for the nuclear hyperspin invariance.  
In absence of the interchain hopping $t_{\perp}$, model (\ref{eqn:hamlatticeladder}) enjoys an U(1)$_{\text{o}}$ orbital  or relative-charge symmetry ($c_{1\alpha,i} \rightarrow e^{i \theta} c_{1\alpha,i}$, $c_{2 \alpha,i} \rightarrow e^{-i \theta} c_{2\alpha,i}$). The latter symmetry is explicitly broken for a non-zero $t_{\perp}$ and the internal symmetry group of model (\ref{eqn:hamlatticeladder}) reduces to:
\begin{equation}
{\rm G} =  {\rm U}(1)_{\text{c}}  \times {\rm SU}(N) \times \mathbb{Z}_{2,\text{o}},
\label{eqn:symhamperp}
\end{equation}
$\mathbb{Z}_{2,\text{o}}$ being the permutation symmetry between the two legs of the ladder.

\subsection{Order parameters and density-current duality}
\label{sec:orderparaduality}

We now introduce several lattice order parameters that will play a central role in our study. In this respect,
we define the band-basis of the lattice fermions:
\begin{equation}
d_{1\alpha,i}=\frac{1}{\sqrt{2}}\left(c_{1 \alpha,i}-c_{2 \alpha,i}\right), d_{2 \alpha,i}=
\frac{1}{\sqrt{2}}\left(c_{1 \alpha,i}+c_{2 \alpha,i}\right) .
\label{cend}
\end{equation}
 
A first order parameter describes an OAF$_{\perp}$ phase  or an orbital-density wave (ODW) along the $y$-axis: 
 \begin{equation}
 \begin{split}
 {\cal O}_{{\rm OAF}_{\perp}} (i)  & = \frac{i}{2} \left(-  c_{1\alpha,i}^\dagger c_{2 \alpha,i} +c_{2 \alpha,i}^\dagger c_{1 \alpha,i} \right) \\
& = \frac{i}{2} \left(-  d_{1\alpha,i}^\dagger d_{2 \alpha,i} +d_{2 \alpha,i}^\dagger d_{1 \alpha,i} \right).
\end{split}
\label{ODWy}
\end{equation}
Such an order parameter describes the formation of a phase where a current circulates along the leg or vertical links of the ladder (see Fig. \ref{figOAFphases}). Similarly, a current circulating along the diagonals of the ladder can also be introduced by the following order parameter (see Fig. \ref{figOAFphases}):
\begin{equation}
\begin{split}
{\cal O}_{{\rm OAF}_{\text d}}  (i)  &=  - \frac{i}{4} \left(c_{1\alpha,i+1}^\dagger c_{2 \alpha,i} - c_{2\alpha,i+1}^\dagger c_{1 \alpha,i} - {\rm H.c.} \right) \\
& =  - \frac{i}{4} \left(d_{1\alpha,i+1}^\dagger d_{2 \alpha,i} - d_{2 \alpha,i+1}^\dagger d_{1 \alpha,i}   
- {\rm H.c.} \right).
\end{split}
\label{diagcurrent}
\end{equation}

On top of these currents, we consider density and bond-ordering operators. The CDW$_{-}$ order parameter is defined by
\begin{equation}
\begin{split}
{\cal O}_{{\rm CDW}_-}  (i)   &=  \frac{1}{2} \left( c_{1\alpha,i}^\dagger c_{1 \alpha,i} - c_{2\alpha,i}^\dagger c_{2 \alpha,i}
\right)  \\
& =  \frac{1}{2} \left( d_{1\alpha,i}^\dagger d_{2 \alpha,i} + d_{2 \alpha,i}^\dagger d_{1 \alpha,i} \right) ,
\end{split}
\label{CDW-}
\end{equation}
which can also be interpreted as an ODW along the $z$-axis. The longitudinal BDW$_{-}$ phase is described
by the following order parameter:
\begin{equation}
 \begin{split}
 {\cal O}_{{\rm BDW}_-}  (i)  & =  \frac{1}{4} \left( c_{1\alpha,i}^\dagger c_{1\alpha,i+1} 
 - c_{2 \alpha,i}^\dagger c_{2 \alpha,i+1} + {\text H.c.}  \right)  \\
& =   \frac{1}{4} \left( d_{1\alpha,i}^\dagger d_{2 \alpha,i+1} +d_{2 \alpha,i}^\dagger d_{1 \alpha,i+1} + {\text H.c.}  \right) .
\end{split}
\label{BDW-para}
\end{equation}
It describes the formation of a bond-dimerization phase which is out of phase between the legs of the ladder (see
Fig. \ref{figcompetingorderslatt}). The last order parameter corresponds to the stabilization of 
a transverse BDW phase or an ODW phase along the $x$-axis:
\begin{equation}
\begin{split}
{\cal O}_{{\rm BDW}_{\perp}}  (i)   &= \frac{1}{2} \left(  c_{1\alpha,i}^\dagger c_{2 \alpha,i} + c_{2\alpha,i}^\dagger c_{1 \alpha,i} \right) \\
&=  - \frac{1}{2} \left(d_{1\alpha,i}^\dagger d_{1 \alpha,i} - d_{2 \alpha,i}^\dagger d_{2 \alpha,i}   \right).
\end{split}
\label{CDWODDxlattice}
\end{equation}

These order parameters are related between themselves by a density-current duality ${\cal D}_{\rm dc}$  symmetry, introduced in the $N=2$ case in Refs. \onlinecite{Momoi-H-03,Momoi-H-05}. This duality symmetry extends to the SU($N$) case and can be defined  through a canonical transformation $U$ on the lattice fermions $c_{l\alpha,i}$ as:
  \begin{equation}
 \begin{split}
 & {\tilde c}_{1 \alpha,i}  = U c_{1 \alpha,i} U^{\dagger} = \frac{1}{\sqrt{2}} \left( e^{i \pi/4} c_{1 \alpha,i}
 + e^{-i \pi/4} c_{2 \alpha,i} \right), \\
& {\tilde c}_{2 \alpha,i}  = U c_{2 \alpha,i} U^{\dagger} = \frac{1}{\sqrt{2}} \left( e^{-i \pi/4} c_{1 \alpha,i}
 + e^{i \pi/4} c_{2 \alpha,i} \right) .
 \end{split}
 \label{dualitylatticefer}
 \end{equation} 
 The transformation is very simple in terms of the band lattice fermions $d_{l \alpha}$: 
 \begin{equation}
 \begin{split}
 &  U d_{1 \alpha,i} U^{\dagger} = i  \; d_{1 \alpha,i} ,
  \\
& U d_{2 \alpha,i} U^{\dagger} =  d_{2 \alpha,i}, \\
& U = \prod_{i, \alpha} e^{- i \frac{\pi}{2} d_{1 \alpha,i}^{\dagger} d_{1 \alpha,i}} .
 \end{split}
 \label{dualitylatticedferd}
 \end{equation} 
 The transformation is canonical and leaves invariant the non-interacting Hamiltonian (\ref{freehamlatt}). In particular, 
 this duality symmetry is valid for all fillings.  Its main interest is to relate a pair of order parameters (\ref{ODWy},\ref{CDW-}) and (\ref{diagcurrent}, \ref{BDW-para}):
  \begin{equation}
 \begin{split}
  U  {\cal O}_{{\rm CDW}_{-}}  U^{\dagger}  & =  {\cal O}_{{\rm OAF}_{\perp}},   \\
  U  {\cal O}_{{\rm OAF}_{\perp}} U^{\dagger}  & = - {\cal O}_{{\rm CDW}_{-}}, \\
  U  {\cal O}_{{\rm BDW}_-} U^{\dagger}  & =  {\cal O}_{{\rm OAF}_{\text d}} ,  \\
  U  {\cal O}_{{\rm OAF}_{\text d}} U^{\dagger}  & = - {\cal O}_{{\rm BDW}_-} ,
 \end{split}
 \label{dualityCDWOAF}
 \end{equation} 
 whereas the density $n_i$ and ${\cal O}_{{\rm BDW}_{\perp}} $ remain invariant under this transformation.
 The duality (\ref{dualitylatticefer}) exchanges thus a density into a current and vice-versa.  The CDW$_{-}$ (respectively 
BDW$_{-}$)  order parameter becomes a current which circulates along the rungs (respectively diagonals) of the ladder
and vice-versa (see Fig. \ref{figcompetingorderslatt}).
 
 The duality symmetry (\ref{dualitylatticefer}) can be enlarged to a U(1) symmetry by introducing
 the following canonical transformation $U(\theta)$ which depends on 
 an angle $\theta$:
 \begin{equation}
 \begin{split}
 & {\tilde c}_{1 \alpha,i}  = U(\theta) c_{1 \alpha,i} U^{\dagger} (\theta)= e^{i \frac{\theta}{2}} \left( c_{1 \alpha,i}
 \cos \frac{\theta}{2} - i  \; c_{2 \alpha,i}   \sin \frac{\theta}{2}\right), \\
& {\tilde c}_{2 \alpha,i}  = U(\theta) c_{2 \alpha,i} U^{\dagger}(\theta) =
 e^{i \frac{\theta}{2}} \left( c_{2 \alpha,i}
 \cos \frac{\theta}{2} - i  \; c_{1 \alpha,i}   \sin \frac{\theta}{2}\right),
 \end{split}
 \label{dualityUthetalatticefer}
 \end{equation} 
 and $U(\pi/2) = U$.
 
Under this transformation, the order parameters, related by the duality symmetry ${\cal D}_{\rm dc}$, transform as a doublet:
\begin{align}
    U(\theta)  \begin{pmatrix}
         {\cal O}_{{\rm OAF}_{\perp}}  \\
         \; \;  {\cal O}_{{\rm CDW}_{-}} 
    \end{pmatrix}   
    U^{\dagger} (\theta)
= \begin{pmatrix}
        \cos \theta & - \sin \theta \\
        \sin \theta & \cos \theta
    \end{pmatrix} 
    \begin{pmatrix}
         {\cal O}_{{\rm OAF}_{\perp}}  \\
       \; \;   {\cal O}_{{\rm CDW}_{-}} 
    \end{pmatrix}   ,
\label{doubletorderpara}
    \end{align}
 and
   \begin{align}
    U(\theta)  \begin{pmatrix}
         {\cal O}_{{\rm OAF}_{\text d}}  \\
         \; \;  {\cal O}_{{\rm BDW}_-} 
    \end{pmatrix}   
    U^{\dagger} (\theta)
= \begin{pmatrix}
        \cos \theta & - \sin \theta \\
        \sin \theta & \cos \theta
    \end{pmatrix} 
    \begin{pmatrix}
         {\cal O}_{{\rm OAF}_{\text d}} \\
       \; \;   {\cal O}_{{\rm BDW}_-} 
    \end{pmatrix}   .
\label{doubletorderpara2}
    \end{align}

The interacting Hamiltonian (\ref{doublewellint}) is stable under the transformation (\ref{dualitylatticefer}) and 
we find the following identification of the coupling constants with obvious notations:
\begin{equation}
 \begin{split}
& {\tilde U} = \frac{1}{2} \left( U + V - \frac{J_{\perp} (N +1)}{2N} - 2 V_{\rm pair} \right), \\
& {\tilde V} = \frac{U(N-1)}{2N}  + \frac{V(N+1)}{2N}  + \frac{J_{\perp} (N^2 -1)}{4N^2} 
+ \frac{V_{\rm pair} (N -1)}{N} , \\
& {\tilde J}_{\perp} = - U + V + \frac{J_{\perp} (N -1)}{2N} - 2 V_{\rm pair} ,  \\
& {\tilde V}_{\rm pair} = - \frac{U}{4} + \frac{V}{4} - \frac{J_{\perp} (N +1)}{8N} +
\frac{V_{\rm pair}}{2} .
\end{split}
 \label{couplingrelation}
\end{equation}
The self-dual space which remains invariant
under the transformation (\ref{dualitylatticefer}) is:
\begin{equation}
U = V - \frac{J_{\perp} (N +1)}{2N} - 2 V_{\rm pair} .
\label{selfdualine}
\end{equation}
The Hamiltonian along this self-dual space is invariant under the U(1) symmetry (\ref{dualityUthetalatticefer}). 
The quantum phase transition between two dual phases is governed by the latter symmetry which rotates a pair of dual order parameters (\ref{doubletorderpara}, \ref{doubletorderpara2}).

The SU($N$) Hubbard two-leg ladder (\ref{hubbard2leg}) is not invariant under the transformation (\ref{couplingrelation}) but  is mapped onto the interacting Hamiltonian (\ref{generalint}) with fine-tuned coupling constants:
\begin{equation}
 \begin{split}
& {\rm H}^{{\cal D}_{\rm dc}}_{\rm Hubbard} = 
  {\rm H}_0  + \frac{U}{4}\sum_{i} \sum_{l=1}^{2} \left(\sum_{\alpha \neq \beta}n_{l  \alpha, i}n_{l  \beta,i} +   2 n_{1,i} n_{2,i} \right) \\
 &  + \frac{U}{2}  \sum_i   \left( c_{1 \alpha,i}^{\dagger} c_{2 \beta,i}^{\dagger}  c_{1 \beta,i}  c_{2 \alpha,i} 
 - \frac{1}{2}   \left( c_{1 \alpha,i}^{\dagger} 
c_{1 \beta,i}^{\dagger}  c_{2 \beta,i}  c_{2 \alpha,i}  \right.  \right. \\
& \left. \left.   + {\rm H.c.} \right) \right) ,
\end{split}
 \label{dualHubbard}
\end{equation}
which has a similar structure as the cold-atom model loaded into a double-well optical lattice (\ref{doubleintfinetuned})
if we perform the transformation $ c_{1 \alpha,i} \rightarrow  i c_{1 \alpha,i}$. Similarly, the dual version of  
the double-well model (\ref{doubleintfinetuned}) is given by:
 \begin{equation}
 \begin{split}
&  {\rm H}^{{\cal D}_{\rm dc}}_{\rm dw} =   {\rm H}_0 +
\frac{U+V}{4}\sum_{i} \sum_{l=1}^{2}\sum_{\alpha \neq \beta}n_{l  \alpha, i}n_{l  \beta,i} \\
& + \frac{U+V}{2}\sum_{i} n_{1,i} n_{2,i} 
  + \frac{U+V}{2}  \sum_i    c_{1 \alpha,i}^{\dagger} c_{2 \beta,i}^{\dagger}  c_{1 \beta,i}  c_{2 \alpha,i}  \\
 & + \frac{3 V - U}{4}  \sum_i   \left( c_{1 \alpha,i}^{\dagger} c_{1 \beta,i}^{\dagger}  c_{2 \beta,i}  c_{2 \alpha,i}   + {\rm H.c.} \right)  ,
\end{split}
 \label{dualdw}
\end{equation}
which includes model (\ref{dualHubbard}) when $V=0$.

\section{Low-energy description at half-filling}
\label{sec:low-energyhalfilling}
In this section, we investigate the low-energy description of model (\ref{eqn:hamlatticeladder}) in the weak-coupling
limit at half-filling by performing a continuum limit complemented by a one-loop RG approach. The main emphasis is laid on the formation of loop-current long-range ordered phases described by the order parameters (\ref{ODWy}) 
and (\ref{diagcurrent}) in some special regions of the full parameter space of model (\ref{eqn:hamlatticeladder}).

\subsection{Continuum limit}
\label{sec:continuum_limit}

In the band-basis (\ref{cend}),  the non-interacting Hamiltonian (\ref{freehamlatt}) is diagonal in the momentum space: 
\begin{equation} 
{\rm H}_0 = \sum_{k} \sum_{l=1}^{2} \left( \epsilon_l(k) - \mu \right)  d^{\dagger}_{l\alpha,k} d_{l\alpha,k} ,
\label{noninterhamlim}
\end{equation} 
where $\epsilon_l(k) = - 2 t \cos (k a_0) - t_\perp (-1)^l$ are the energy-bands of the ladder and 
$a_0$ is the lattice spacing. In  the following, we restrict our analysis to the case where we have 
four distinct Fermi points $\pm k_{1,2F}$. At half-filling, one has $k_{1F} + k_{2F} = \pi /a_0$ and the anisotropy $\delta = a_0 (k_{2F} - k_{1F}) $ satisfies $\sin( \delta/2) = t_{\perp}/(2t) $ with $t_{\perp} < 2 t $. In the low-energy limit,  the spectrum of the non-interacting Hamiltonian (\ref{noninterhamlim}) may be linearized around 
these Fermi points with a unique Fermi velocity $v_{F}= v_{1F} = 2 t \sin (k_{1F} a_0) =  2 t \sin (k_{2F} a_0) = v_{2F}$. 

The starting point of the analysis is  the continuum description of the lattice fermionic operators
$d_{l\alpha,\,i}$ in terms of $2N$ left-right moving Dirac fermions 
($l=1,2$, $\alpha=1,\ldots,N$): \cite{Gogolin-N-T-book,Giamarchi}
\begin{equation}
d_{l \alpha,\,i} \rightarrow \sqrt{a_0} \left(L_{l \alpha}(x)
e^{-i k_{l F} x} + R_{l \alpha}(x) e^{i k_{l F} x} \right),
\label{contlimitDirac}
\end{equation}
with  $x= i a_0$.  The non-interacting Hamiltonian density is equivalent to that of $2N$ left-right moving Dirac fermions:
\begin{equation}
  {\cal H}_0=-i v_{F}    \left(:R_{l \alpha} ^\dag \partial_x R_{l \alpha} ^{\phantom \dag}: - 
  : L_{l \alpha}^\dag \partial_x L_{l \alpha}^{\phantom \dag}: \right) ,
\label{HamcontDirac}  
\end{equation}  
where now the summation over all repeated indexes is implied in the following.  In Eq. (\ref{HamcontDirac}), a normal ordering with respect to the Fermi seas denoted by $::$ is assumed.

The  lattice model (\ref{eqn:hamlatticeladder}) has important discrete symmetries which act on the Dirac fermions (\ref{contlimitDirac}) of the continuum limit. The one-step translation invariance T$_{a_0}$ is described by: 
\begin{equation}
L_{l \alpha} \xrightarrow{{\rm T}_{a_0}} e^{- i k_{l F} a_0} L_{l \alpha}, \; \;  R_{l \alpha} \xrightarrow{{\rm T}_{a_0}} e^{i k_{l F} a_0} R_{l \alpha} .
\label{Ta0}  
\end{equation}  
The effect of the $\mathbb{Z}_{2,\text{o}}$ symmetry ($1\leftrightarrow 2$) and site-parity P$_s$ ($c_{l \alpha,i} \rightarrow  c_{l \alpha,-i}$) on the  Dirac fermions read respectively as follows:
\begin{equation}
L_{l \alpha} \xrightarrow{\mathbb{Z}_{2,\text{o}}} (-)^{l} L_{l \alpha}, \; \;  R_{l \alpha}  \xrightarrow{\mathbb{Z}_{2,\text{o}}}  (-)^{l} R_{l \alpha} ,
\label{Z2o}  
\end{equation}  
\begin{equation}
L_{l \alpha} (x)   \xrightarrow{{\rm P}_{\rm s}}   R_{l \alpha} (-x), \; \;  R_{l \alpha} (x)  \xrightarrow{{\rm P}_{\rm s}}   L_{l \alpha} (-x) .
\label{Ps}  
\end{equation}  
The link parity P$_{L}$ ($c_{l \alpha,i} \rightarrow  c_{l \alpha,1-i}$) is a combination of the site parity and the one-step translation symmetry.

On top of these discrete symmetries, the non-interacting model \eqref{HamcontDirac} displays an U($N$)$|_\text{L}$ $\times$ U($N$)$|_\text{R}$ continuous symmetry for each chain $l=1,2$. This symmetry results from its invariance under independent unitary transformations on the $2N$ left and right Dirac fermions. The gapless properties of the $2N$ Dirac fermionic model (\ref{HamcontDirac}) are in turn described by a conformal field theory (CFT) based on the U($N$)  group. In this respect, we introduce the U(1) and SU($N$)$_1$ chiral currents for each chain $l=1,2$: 
\begin{equation}
\begin{split}
& J_{lL} =  :L_{l \alpha}^\dagger  L_{l\alpha}: ,\\
& J_{lL}^A = L_{l \alpha}^\dagger T^A_{\alpha\beta} L_{ l\beta}   ,
\end{split}
\label{currentsaniso}
\end{equation}
with similar definitions for the right chiral currents. The U(1) (respectively SU($N$)$_1$)  $J_{lL,R}$ (respectively
$J_{lL,R}^A$) currents describe a U(1) (respectively SU($N$)$_1$) CFT with central charge $c=1$ (respectively $c=N-1$). 
The non-interacting massless Dirac Hamiltonian density  (\ref{HamcontDirac}) can be expressed in terms of these currents from the so-called Sugawara construction: \cite{Affleck-NP86,Affleck-88,Gogolin-N-T-book,James-K-L-R-T-18} 
\begin{equation}
  {\cal H}_0=\sum_{l=1}^{2} \frac{\pi v_{F}}{N} : J_{lL}^2: + 
   \frac{2\pi v_{F}}{N+1}  : J_{lL}^A J_{lL}^A: + L \rightarrow R  ,
\label{HamcontDiracurrents}  
\end{equation}  
which describes a metallic phase C2S$2N-2$, where C$n$S$m$ denotes a gapless phase with $n$ gapless charge modes and $m$ gapless spin modes \cite{Balents-F-96}. According to Eq. (\ref{eqn:symhamperp}),
the lattice model (\ref{eqn:hamlatticeladder}) has only an U($N$) = U(1) $\times$ SU($N$) continuous symmetry. 
It is thus a useful to introduce a single U(1)  current  $J_{c L,R} = J_{1L,R}  + J_{2L,R} $ which describes 
the U(1) charge symmetry of model (\ref{eqn:hamlatticeladder}). Similarly, the SU($N$) nuclear spin degrees of freedom are captured by an SU($N$)$_2$ CFT generated by the symmetric combination of the two SU($N$)$_1$ currents:  $J_{L,R}^A = J_{1L,R}^A + J_{2L,R}^A$. The latter CFT has a central charge $c= 2(N^2 - 1)/(N+2)$ \cite{DiFrancesco-M-S-book}.

On top of the currents (\ref{currentsaniso}) related to the existence of an SU($N$) symmetry, we need to consider other fermionic bilinears to describe the continuum limit  of the lattice model. In particular, we introduce orbital SU(2)$_N$ chiral currents: 
\begin{equation}
 j_{L}^i = \frac{1}{2} L_{m\alpha}^\dagger \sigma^i_{mn} L_{n\alpha} , 
 j_{R}^i = \frac{1}{2} R_{m\alpha}^\dagger \sigma^i_{mn} R_{n\alpha},
\label{orbitalcurrents}
\end{equation}
where $\sigma^i$ ($i=x,y,z$) are the standard Pauli matrices. The $z$-component of this current describes the conservation of the relative number of atoms between the legs of the ladder, i.e., the relative-charge degrees of freedom: 
$2 j_{L,R}^z = J_{1L,R}  - J_{2 L,R} $. We also consider fermionic bilinears  which couple SU($N$) spin and orbital degrees of freedom:
\begin{equation}
J_{L}^{A,i} = L_{m\alpha}^\dagger T^{A,i}_{m,\alpha;n,\beta} L_{n\beta}, 
J_{R}^{A,i} = R_{m\alpha}^\dagger T^{A,i}_{m,\alpha;n,\beta} R_{n\beta},
\label{spinorbitalcurrents}
\end{equation}
with $T^{A,i} = \frac{1}{\sqrt{2}}T^A\otimes \sigma^i $.  

The set of left  and right currents 
\begin{equation}
{\rm U}(2N)_1 \; {\rm currents} : \{ J_{c L,R} , J_{L,R}^A,  j_{L,R}^i, J_{L,R}^{A,i} \},
\label{u2N}
\end{equation}
defines an U($2N$)$_1$ current which generates the U($2N$)$_1$  CFT  with central charge $c=2N$. The latter CFT  describes the critical theory of the non-interacting model. The U($2N$)$_1$ current (\ref{u2N}) contains an SU($2N$)$_1$ current $\mathcal{J}_{L,R}^{A}$ ($A= 1, \ldots, 4N^2-1$) which can be defined as follows:
\begin{equation}
\begin{split}
\mathcal{J}_{L}^{A} & =  L_{a}^{\dagger} {\cal T}^A_{ab}  L_{b} ,\\
\mathcal{J}_{R}^{A} & =  R_{a}^{\dagger} {\cal T}^A_{ab}  R_{b},
\end{split}
\label{SU(2N)_1curr}
\end{equation}
 where $a= (l, \alpha)$ and  $b= (m, \beta)$ run from $1$ to $2N$, and ${\cal T}^A$  are SU($2N$) generators in the fundamental representation of the SU(2$N$) group with: ${\rm Tr}( {\cal T}^A {\cal T}^B) = \delta^{AB}/2$. The free-Hamiltonian densities (\ref{HamcontDirac}) and (\ref{HamcontDiracurrents}) can then be expressed in terms of the U($2N$)$_1$ current (\ref{u2N}):
\begin{equation}
  {\cal H}_0= \frac{\pi v_{F}}{2N} : J_{cL}^2: + 
   \frac{2\pi v_{F}}{2N+1}  : \mathcal{J}_{L}^{A} \mathcal{J}_{L}^{A}: + L \rightarrow R  .
\label{HamcontDiracurrentsU(2N)}  
\end{equation}  

Finally, we need additional operators that describe the umklapp processes at half-filling since $k_{1F} + k_{2F} = \pi/a_0$: 
\begin{equation}
\begin{split}
& A^{\alpha \beta +}_{mn L} =
\frac{-i}{2} \left(  L^{\dagger}_{m \alpha} L^{\dagger}_{n \beta} 
- L^{\dagger}_{m \beta}  L^{\dagger}_{n \alpha}  \right) , \\
& A^{\alpha \beta -}_{mn L} =
\frac{-i}{2} \left(  L_{m \alpha} L_{n \beta} 
- L_{m \beta}  L_{n \alpha}  \right), \\
& S^{\alpha \beta +}_{L} =  \frac{1}{2} \left(  L^{\dagger}_{1 \alpha} L^{\dagger}_{2 \beta} 
+ L^{\dagger}_{1 \beta}  L^{\dagger}_{2 \alpha}  \right), \\
& S^{\alpha \beta -}_{L} = - \frac{1}{2} \left(  L_{1 \alpha} L_{2 \beta} 
+ L_{1 \beta}  L_{2 \alpha}  \right),
\end{split}
\label{umklappop}
\end{equation}
with a similar set of operators for the right fields.

With all these definitions at hands, one can perform the continuum limit of the lattice interacting Hamiltonian
(\ref{doublewellint}) at half-filling. It is invariant under the U(1)$_{\text{o}}$ orbital symmetry ($c_{1\alpha,i} \rightarrow e^{i \theta} c_{1\alpha,i}$, $c_{2 \alpha,i} \rightarrow e^{-i \theta} c_{2\alpha,i}$). In this respect, it is useful 
to  introduce a chiral transformation $\Omega$ on the left-moving Dirac fermions to make explicit this U(1)$_{\text{o}}$  invariance: 
\begin{equation}
\Omega: \;  L_{1 \alpha}  \xrightarrow{\Omega}   i L_{2 \alpha} , \; \; 
 L_{2 \alpha}  \xrightarrow{\Omega}  - i L_{1 \alpha} ,
\label{chiraltrans}
\end{equation}
whereas the right-moving fermions remain unchanged.

The continuum limit of the interacting Hamiltonian (\ref{doublewellint})  after
performing the  chiral transformation $\Omega$ (\ref{chiraltrans})  is given by:
\begin{equation}
\begin{split}
& \mathcal{H}^{\Omega}_{\text{int}} \\
&= g_1 J_{L}^A J_{R}^A + g_2 \left(
J_{L}^{A,x} J_{R}^{A,x} + J_{L}^{A,y} J_{R}^{A,y} \right) + g_3 J_{L}^{A,z} J_{R}^{A,z}  \\
  & + g_4 \left(j_{L}^x j_{R}^x + j_{L}^y j_{R}^y \right) 
  +g_5 j_{L}^zj_{R}^z + g_6 J_{c L} J_{c R}  \\
  & + g_7 \left(S^{\alpha \beta +}_{L}  S^{\alpha \beta -}_{R} +\textrm{H.c.}\right) 
  + \frac{g_8}{2} \sum_{m=1,2} \left(A^{\alpha \beta +}_{mm L}  
  A^{\alpha \beta -}_{mm R} +\textrm{H.c.}\right) \\
 &+ g_9  \left(A^{\alpha \beta +}_{1 2 L}  A^{\alpha \beta -}_{1 2 R} +\textrm{H.c.}\right) ,
\end{split}
  \label{lowenergyhamhalfilling}
\end{equation}
with an explicit U(1)$_{\text o}$ invariance along the $z$-axis in the orbital space thanks to 
the $\Omega$ transformation (\ref{chiraltrans}). The initial conditions of the RG flow are:
\begin{equation}
\begin{split}
& g_1 =   \left( - U + \frac{J_{\perp}}{2} \right)a_0,  \\
& g_2 =   \left(U - V + \frac{(N+1)J_{\perp}}{2N} + 2 V_{\rm pair}\right)a_0 , \\
& g_3 =   \left(2 V - \frac{J_{\perp}}{N} + 4 V_{\rm pair}\right) a_0 ,  \\
& g_4 = \frac{N-1}{N} \left(V  - U  - \frac{(N+1)J_{\perp}}{2N} - 2 V_{\rm pair} \right) a_0 , \\
& g_5 =   \left( \frac{2 V}{N}  + \frac{(N^2 - 1)J_{\perp}}{N^2} - \frac{4(N-1)V_{\rm pair} }{N}  \right) a_0 , \\
& g_6 = \left(\frac{(N - 1) U}{2N} + \frac{V}{2} \right) a_0 , \\
& g_7 =   - \left( V + \frac{(N - 1)J_{\perp}}{2N} \right) a_0 ,  \\
& g_8 =  - \left( \frac{U}{2} - \frac{V}{2}   + \frac{(N+1)J_{\perp}}{4N} + V_{\rm pair} \right) a_0,  \\
& g_9 =  \left(U  - 2 V_{\rm pair} \right)a_0  .
\end{split}
\label{couplingshalfilling}
\end{equation}
In the derivation of the continuum limit (\ref{lowenergyhamhalfilling}), 
we have used that $k_{1F} \ne k_{2F}$  and all chiral contributions have been neglected as usual.

\subsection{Duality approach to competing ordered phases}
\label{sec:1loopRGhalfilling}

The field theory  (\ref{lowenergyhamhalfilling}) is stable under the RG approach. 
The one-loop RG equations for  this model  have been derived in Ref. \onlinecite{Bois-C-L-M-T-15} and they are given by:
\begin{equation}
\begin{split}
\dot{g_1} =& \frac{N}{4\pi} g^2_1 + \frac{N}{8\pi} g^2_2 + \frac{N}{16\pi} g^2_3
  + \frac{N+2}{4\pi} g^2_7 \\
  &+ \frac{N-2}{4\pi} \left( 2 g^2_8 + g^2_9  \right) , \\
\dot{g_2} =&  \frac{N}{2\pi} g_1 g_2 + 
    \frac{N^2 -4}{4\pi N} g_2 g_3 +   \frac{1}{2\pi} (g_2 g_5+g_3 g_4)   \\
&+ \frac{N}{\pi} g_7 g_8 +  \frac{N-2}{\pi} g_8 g_9 ,  \\
\dot{g_3} =&  \frac{N}{2\pi} g_1 g_3 +  
    \frac{N^2 -4}{4\pi N} g^2_2 +    \frac{1}{\pi} g_2 g_4 
     + \frac{N}{\pi} g_7 g_9 +  \frac{N-2}{\pi} g^2_8 , \\
\dot{g_4} =&  \frac{1}{2\pi} g_4 g_5 + 
    \frac{N^2 - 1}{2\pi N^2} g_2 g_3 +    \frac{2(N-1)}{\pi N} g_8 g_9 ,  \\
 \dot{g_5} =& \frac{N^2 -1}{2\pi N^2} g^2_2 + \frac{1}{2\pi} g^2_4 + \frac{2(N-1)}{\pi N} g^2_8,   \\
\dot{g_6} =& \frac{N +1}{4\pi N} g^2_7 +  \frac{N - 1}{2\pi N} g^2_8 + \frac{N-1}{4\pi N} g^2_9,  \\
\dot{g_7} =&  \frac{(N + 2)(N - 1)}{2\pi N} g_1 g_7 +  \frac{2}{\pi} g_6 g_7 \\
& +    \frac{N-1}{4\pi} (2 g_2 g_8 + g_3 g_9), \\
\dot{g_8} =&  \frac{N + 1 }{4\pi} g_2 g_7 +  \frac{2}{\pi} g_6 g_8  
+  \frac{1}{2\pi} \left(g_4 g_9 + g_5 g_8 \right) \\
& +  \frac{(N-2)(N+1)}{4\pi N} \left(2 g_1 g_8 + g_2 g_9 + g_3 g_8 \right), \\
\dot{g_9} =&  \frac{N + 1 }{4\pi} g_3 g_7 +  \frac{1}{\pi} (g_4 g_8 + 2 g_6 g_9 ) \\
&+  \frac{(N-2)(N+1)}{2\pi N} \left( g_1 g_9 + g_2 g_8 \right),
\end{split}
\label{RGhalf}
\end{equation}
where ${\dot g}_i = \partial g_i/ \partial l (i =1,\ldots, 9)$ with $l$ being the RG time.  It is useful to introduce the
rescaling variables $f_i$:
\begin{equation}
\begin{split}
& f_{1,7,8,9} = \frac{N}{\pi} g_{1,7,8,9}  \, , \;\;   f_{2,3} = \frac{N}{2 \pi} g_{2,3} ,\\
&  f_{4,5} = \frac{N^2}{2 \pi} g_{4,5}  \, , \;\;  f_{6} = \frac{2 N^2}{\pi} g_{6} . 
\end{split}
 \label{rescaling}
 \end{equation}
  
 The phases of the lattice model (\ref{eqn:hamlatticeladder}) can be obtained  by solving numerically  these RG equations
 and by identifing specific RG rays in the far infrared (IR) limit which lead to the determination of the weak-coupling phases.
 This approach will be discussed in Sec.  \ref{sec:1loopphasediaghalfilled}. Here, our goal is to investigate the possible 
 stabilization of the four competing orders ${\cal O}_{{\rm OAF}_{\perp}}$ (\ref{ODWy}), ${\cal O}_{{\rm OAF}_{\text d}}$ (\ref{diagcurrent}), ${\cal O}_{{\rm CDW}_{-}} $ (\ref{CDW-}),  and ${\cal O}_{{\rm BDW}_{-}}$ (\ref{BDW-para}) by considering  the stable fully symmetric line  $f_i = f$ in model (\ref{lowenergyhamhalfilling}) and exploiting the existence of 
 non-perturbative hidden duality symmetries.
 
\subsubsection{SO(4N) Gross-Neveu model}
\label{sec:GN}

Along the $f_i = f$ line, model (\ref{lowenergyhamhalfilling}) reduces to the interacting Hamiltonian density of the SO($4N$)  Gross-Neveu (GN) model: \cite{Gross-N-74}
\begin{equation}
\begin{split}
{\cal H}_{\text{GN}} =& -i v_\text{F}\left(:R_{l \alpha} ^\dag \partial_x R_{l \alpha} ^{\phantom \dag}: - 
  :L_{l \alpha}^\dag \partial_x L_{l \alpha}^{\phantom \dag}: \right) \\
  &+ \frac{\pi f}{2N} \left( L_{l \alpha} ^{\dag}  R_{l \alpha} - \text{H.c.} \right)^2,
  \end{split}
\label{GNso}
\end{equation}
where the SO($4N$)  symmetry stems from the decomposition of $2N$ Dirac fermions into $4N$ Majorana (real) 
fermions: 
\begin{equation}
\begin{split}
L_{l \alpha} & = ( \xi_{l \alpha L} + i \chi_{l \alpha L})/\sqrt{2},\\
R_{l \alpha} & = ( \xi_{l \alpha R} + i \chi_{l \alpha R})/\sqrt{2} .
\end{split}
\label{majo}
\end{equation}
By combining these Majorana fermions into a single $4N$-component Majorana field: ${\vec \Psi}_{L,R} = ( \xi_{1 \alpha L,R} , \chi_{1 \alpha L,R}, \xi_{2 \alpha L,R} 
, \chi_{2 \alpha L,R})^{T}$, the SO($4N$)  symmetry of  the GN model (\ref{GNso}) becomes more explicit: 
\begin{equation}
\begin{split}
 {\cal H}_{\text{GN}} & =  
   - \frac{i v_\text{F}}{2}  \left( :{\vec \Psi_{R}} \cdot  \partial_x  {\vec \Psi_{R}}:   - 
  :{\vec \Psi_{L}} \cdot  \partial_x  {\vec \Psi_{L}}: \right) \\
 &  + \frac{\pi f}{2N} \left( {\vec \Psi_{L}} \cdot {\vec \Psi_{R}} \right)^2.
 \end{split}
\label{GNmajo}
\end{equation}
This SO($4N$) symmetry is the maximal continuous symmetry enjoyed by the $2N$ Dirac fermions. 

The main interest of the $f_i = f$ line stems from the fact that the SO($4N$) GN model is a massive integrable 
field theory when $f>0$ whose mass spectrum is known exactly \cite{Zamolodchikov-Z-79,Karowski-T-81}.
A fully-gapped Mott-insulating phase C0S0 is then stabilized non-perturbatively along this special line. The order parameter of this phase expresses in terms of the Dirac fermions:
\begin{equation}
{\cal O}_{\rm GN} =  i \left( L_{l \alpha} ^{\dag}  R_{l \alpha} - \text{H.c.} \right),
\label{orderparameterGN}
\end{equation}
and condenses, $\langle {\cal O}_{\rm GN} \rangle \ne 0$, in the ground-state of the SO($4N$) GN model (\ref{GNso}).  It turns out that the order parameter (\ref{orderparameterGN}) is directly related to the continuum description of the
diagonal-current order parameter ${\cal O}_{{\rm OAF}_{\text d}}$ of Eq. (\ref{diagcurrent}). Indeed, we first perform
the continuum limit of Eq. (\ref{diagcurrent}) by defining 
\begin{equation}
{\cal O}_{{\rm OAF}_{\text d}} = (-1)^{i} {\cal O}_{{\rm OAF}_{\text d}}(i) ,
\label{contlimitorderpara}
\end{equation} 
and keeping the uniform contribution which involves chiral invariant fields, we get
\begin{equation}
{\cal O}_{{\rm OAF}_{\text d}}  \simeq  \frac{ \sin(k_{1F} a_0)}{2}   \left(L_{1\alpha}^\dagger R_{2 \alpha} 
- L_{2\alpha}^\dagger R_{1 \alpha} + {\rm H.c.} \right).
\label{diagcurrentcont}
\end{equation}
Using the chiral transformation $\Omega$ (\ref{chiraltrans}), the order parameter transforms as:
\begin{equation}
\begin{split}
{\cal O}_{{\rm OAF}_{\rm d}}  \xrightarrow{\Omega}  
{\cal O}^{\Omega}_{{\rm OAF}_{\text d}}  & \simeq  - \frac{i}{2} \sin(k_{1F} a_0)  \left(L_{l\alpha}^\dagger R_{l \alpha} 
- {\rm H.c.} \right)\\
& \simeq  - \frac{\sin(k_{1F} a_0)}{2}   \;  {\cal O}_{\rm GN},
\end{split}
\label{diagcurrentcontdual}
\end{equation}
and thus $\langle {\cal O}^{\Omega}_{{\rm OAF}_{\text d}}  \rangle \ne 0$ in the ground-state of the SO($4N$) GN model (\ref{GNso}). It signals the formation of a fully gapped phase which corresponds to a loop-current ordered phase with
a circulation of currents along the diagonals of the ladder in a staggered way (see Fig. \ref{figOAFphases}): 
$\langle {\cal O}_{{\rm OAF}_{\text d}}  \rangle \ne 0$.  The discrete symmetries of this order parameter can be obtained from its continuous description (\ref{diagcurrentcont}) (see Table \ref{tabsum}).
Using the transformations (\ref{Ta0}), (\ref{Z2o}), 
(\ref{Ps}) on the Dirac fermions, the order parameter ${\cal O}_{{\rm OAF}_{\text d}}$ (\ref{diagcurrentcont}) is odd under
the one-step translation symmetry T$_{a_0}$, the permutation symmetry of the two chains $\mathbb{Z}_{2,\text{o}}$,  the site-parity symmetry P$_{\rm s}$, and even under the link-parity symmetry  P$_{\rm L}$. It is also odd under the time-reversal symmetry T due to the current nature of the lattice order parameter. The Mott-insulating phase is two-fold degenerate with two different ground states which are obtained by changing the orientation of the circulation of the currents of Fig. \ref{figOAFphases}.

\begin{table}[t]
\begin{center}
\begin{ruledtabular}
\begin{tabular}{c|ccccc|ccc} 
& \multicolumn{5}{c}{symmetry} & 
\multicolumn{3}{c}{duality} 
\\
\cline{2-9}
\raisebox{1.5ex}[0pt]{order param.} & 
\makebox{${\rm T}$} & \makebox{${\rm T}_{a_0}$} & 
\makebox{$\mathbb{Z}_{2,\text{o}}$} & 
\makebox{${\rm P}_{\text{s}}$} & 
\makebox{${\rm P}_{\text{L}}$} &
\makebox{${\cal D}$} &
\makebox{${\cal D}_{\rm dc}$} &
\makebox{${\cal D} {\cal D}_{\rm dc}$} \\
\hline
${\cal O}_{{\rm OAF}_{\perp}}$ & $-1$ & $-1$ & $-1$ & $1$ & $-1$ 
& ${\cal O}_{{\rm OAF}_{\text d}}$ 
& ${\cal O}_{{\rm CDW}_{-}}$ 
& ${\cal O}_{{\rm BDW}_{-}} $ \\
\hline 
${\cal O}_{{\rm OAF}_{\text d}}$ & $-1$ & $-1$ & $-1$ & $-1$ & $1$ 
& ${\cal O}_{{\rm OAF}_{\perp}}$ 
& ${\cal O}_{{\rm BDW}_{-}}$ 
& ${\cal O}_{{\rm CDW}_{-}}$ \\
\hline 
${\cal O}_{{\rm CDW}_{-}}$  & $1$ & $-1$ & $-1$ & $1$ & $-1$ 
& ${\cal O}_{{\rm BDW}_{-}}$ 
& ${\cal O}_{{\rm OAF}_{\perp}}$ 
& ${\cal O}_{{\rm OAF}_{\text d}}$ \\
\hline 
${\cal O}_{{\rm BDW}_{-}}$ & $1$ & $-1$ & $-1$ & $-1$ & $1$ 
& ${\cal O}_{{\rm CDW}_{-}}$ 
& ${\cal O}_{{\rm OAF}_{\text d}}$
& ${\cal O}_{{\rm OAF}_{\perp}}$ \\
\end{tabular}
\end{ruledtabular}
\caption{The four order parameters and their discrete and duality symmetries.}
\label{tabsum}
\end{center}
\end{table}

\subsubsection{Duality symmetries}
\label{sec:dualityGN}

On top of the existence of this loop-current ordered phase, the low-energy Hamiltonian (\ref{lowenergyhamhalfilling}) enjoys hidden duality symmetries which correspond to the different $\mathbb{Z}_{2}$-gradings of the Lie algebra $\mathfrak{so}(4N)$ of the 
maximal continuous symmetry of the problem, i.e., SO($4N$) \cite{Boulat-A-L-09}.  A  $\mathbb{Z}_{2}$-grading of 
$\mathfrak{so}(4N)$ is an involutive automorphism $\omega$ such that 
$\mathfrak{so}(4N) = \mathfrak{g}_{\parallel} \oplus  \mathfrak{g}_{\perp}$ with
$\omega(X)=X$ (respectively $\omega(X)=-X$)  $\forall X\in \mathfrak{g}_{\parallel}$ (respectively 
$\mathfrak{g}_{\perp}$). Under the action of $\omega$, one has in a compact form:
\begin{equation} 
\left[ \mathfrak{g}_{\parallel},  \mathfrak{g}_{\parallel} \right] = \mathfrak{g}_{\parallel} , \; 
\left[ \mathfrak{g}_{\parallel},  \mathfrak{g}_{\perp} \right] = \mathfrak{g}_{\perp}, \; 
\left[ \mathfrak{g}_{\perp},  \mathfrak{g}_{\perp} \right] = \mathfrak{g}_{\parallel} .
\label{Z2gradings}
\end{equation}
There exists a complete classification of $\mathbb{Z}_2$-gradings for simple
Lie algebras which are related to the classification of symmetric spaces \cite{hegalson}. For $\mathfrak{so}(4N)$, there are
two classes of  $\mathbb{Z}_2$-gradings: the symmetric classes DIII ($ \mathfrak{g}_{\parallel} =\mathfrak{u} (2N)
$) and BDI ($\mathfrak{g}_{\parallel} = \mathfrak{so}(2N)  \oplus    \mathfrak{so}(2N)$), which give two possible classes of duality symmetries here.

A first duality symmetry, that we note ${\cal D}$,  is described by the DIII class with $ \mathfrak{g}_{\parallel} =  \mathfrak{u} (2N)$. This $\mathbb{Z}_2$-grading has the following action on the left currents of 
the low-energy model (\ref{lowenergyhamhalfilling}):
\begin{equation}
\begin{split}
J^{A}_L &  \xrightarrow{\cal D}  J^{A}_L , J^{A,i}_L  \xrightarrow{\cal D} J^{A,i}_L,  j^{i}_L  
 \xrightarrow{\cal D}   j^{i}_L , 
J_{cL}   \xrightarrow{\cal D}   J_{cL}, \\
S_L^{\alpha \beta +} &  \xrightarrow{\cal D}  - S_L^{\alpha \beta +}, A_{ll L}^{\alpha \beta +}   \xrightarrow{\cal D}  - A_{ll L}^{\alpha \beta +}, 
A_{12 L}^{\alpha \beta +}   \xrightarrow{\cal D}  - A_{12 L}^{\alpha \beta +} ,
\end{split}
\label{currdualitytildeOmega}
\end{equation}
where the invariant part corresponds to the U($2N$)$_1$ currents (\ref{u2N}) of the effective Hamiltonian (\ref{lowenergyhamhalfilling}) since $ \mathfrak{g}_{\parallel} =  \mathfrak{u} (2N)$.  
The line $f_1 = f_2 = f_3 =  f_4 = f_5 = f_6 = -f_7 = - f_8 = -f_9 = f >0$ of model (\ref{lowenergyhamhalfilling}) is then
related to the SO($4N$) GN model (\ref{GNso}) by the duality ${\cal D}$. It signals the formation of a second 
fully-gapped C0S0 Mott-insulating phase which is dual to the OAF$_{\rm d}$ phase. Its physical nature can be deduced by expressing the duality $\cal D$ (\ref{currdualitytildeOmega}) directly in terms of the Dirac fermions of Eq. (\ref{lowenergyhamhalfilling}). In particular, the transformation  (\ref{currdualitytildeOmega})  corresponds to the following chiral transformation on the left-moving Dirac fermions of Eq. (\ref{lowenergyhamhalfilling}):
\begin{equation}
{\cal D}: \;  L_{l \alpha}   \xrightarrow{\cal D}    i \;  L_{l \alpha} ,
\label{OAFduality}
\end{equation}
and the right-ones remain invariant. This duality is highly non-local in terms of the original lattice fermions. In this respect, 
${\cal D}$ is an emergent duality symmetry at low-energy and probably does not exist on the lattice in stark contrast
to ${\cal D}_{\rm dc}$. 

The nature of the second Mott-insulating phase corresponds to the OAF$_{\perp}$ phase described by the order parameter ${\cal O}_{{\rm OAF}_{\perp}}$ (\ref{ODWy}). Indeed, 
its continuous description, defined as in Eq. (\ref{contlimitorderpara}),  is given by: 
\begin{equation}
 {\cal O}_{{\rm OAF}_{\perp}}  \simeq  \frac{i}{2}   \left(  L_{2\alpha}^\dagger R_{1 \alpha} 
  -  L_{1\alpha}^\dagger R_{2 \alpha} - {\rm H.c.} \right) ,
 \label{contOAFperp}
\end{equation}
so that under  the chiral transformation $\Omega$ (\ref{chiraltrans}), we have
\begin{equation}
{\cal O}_{{\rm OAF}_{\perp}}  \xrightarrow{\Omega}  
 {\cal O}^{\Omega}_{{\rm OAF}_{\perp}} = -  \frac{1}{2}  \left(  L_{l\alpha}^\dagger R_{l \alpha} 
  + {\rm H.c.} \right).
\label{OAFtilde}
\end{equation}
Applying then the duality $\cal D$, one has
\begin{equation}
\begin{split}
{\cal O}_{{\rm OAF}_{\perp}}  \xrightarrow{\Omega {\cal D} }    {\cal O}^{\Omega {\cal D} }_{{\rm OAF}_{\perp}}  &=  \frac{i}{2}   \left(  L_{l\alpha}^\dagger R_{l \alpha}  - {\rm H.c.} \right) \\
& = \frac{1}{2}   {\cal O}_{\rm GN} .
\end{split}
\label{OAFdualityGN}
\end{equation}
From the long-range ordering of the order parameter (\ref{orderparameterGN}) of the SO(4$N$) GN model (\ref{GNso}), we deduce the stabilization of the OAF$_{\perp}$ phase with the long-range ordering of the current (\ref{ODWy}):
$\langle {\cal O}_{{\rm OAF}_{\perp}}   \rangle \ne 0$, which is dual to the OAF$_{\rm d}$ phase by the duality 
$\cal D$ (\ref{OAFduality}). 

The transverse currents are accompanied by the alternative alignement of the in-chain currents:
 \begin{equation}
 \begin{split}
 {\cal O}_{{\rm OAF}_{\parallel}} (i) & = \frac{i}{4} \left(c_{1\alpha,i}^\dagger c_{1\alpha,i+1} 
 - c_{2 \alpha,i}^\dagger c_{2 \alpha,i+1} - {\text H.c.} \right) \\
& = \frac{i}{4}  \left(d_{1\alpha,i}^\dagger d_{2 \alpha,i+1} +d_{2 \alpha,i}^\dagger d_{1 \alpha,i+1} - {\rm H.c.} \right).
\end{split}
\label{OAFpara}
\end{equation}
Defining ${\cal O}_{{\rm OAF}_{\parallel}} = (-)^{i} {\cal O}_{{\rm OAF}_{\parallel}} (i)$ as in Eq. (\ref{contlimitorderpara}),
its continuum description in terms of the Dirac fermions reads:
\begin{equation}
\begin{split}
  {\cal O}_{{\rm OAF}_{\parallel}} & \simeq  \frac{i}{2}   \cos(k_{1F} a_0)  \left(  L_{2\alpha}^\dagger R_{1 \alpha} 
  -  L_{1\alpha}^\dagger R_{2 \alpha} - {\rm H.c.} \right) \\
 &  \simeq   \cos(k_{1F} a_0)  {\cal O}_{{\rm OAF}_{\perp}},
 \end{split}
\label{contOAFparallel}
\end{equation}
and thus condenses in the C0S0 OAF$_{\perp}$ phase: $\langle {\cal O}_{{\rm OAF}_{\parallel}}   \rangle \ne 0$.
The phase corresponds to the formation of a loop-current ordering along the plaquette of the ladder (see Fig. \ref{figOAFphases}). 

A second duality corresponds to the density-current duality ${\cal D}_{\rm dc}$ (\ref{dualitylatticefer}).  The relative-charge density phase (\ref{CDW-})  and longitudinal relative BDW operator (\ref{BDW-para}) can be expressed in terms of the left-right moving Dirac fermions of the continuous description (\ref{contlimitDirac}) using similar definitions as in Eq. (\ref{contlimitorderpara}):
\begin{equation}
\begin{split}
 {\cal O}_{{\rm CDW}_{-}} & \simeq  \frac{1}{2}  \left(  L_{1\alpha}^\dagger R_{2 \alpha} 
  + L_{2 \alpha}^\dagger R_{1 \alpha} + {\rm H.c.} \right), \\
{\cal O}_{{\rm BDW}_{-}} & \simeq   \frac{i}{2} \sin(k_{1F} a_0)  \left(L_{1\alpha}^\dagger R_{2 \alpha} 
+ L_{2\alpha}^\dagger R_{1 \alpha} - {\rm H.c.} \right) .
\end{split}
\label{contcompetorderCDWBDW}
\end{equation}
The density-current duality ${\cal D}_{\rm dc}$ (\ref{dualitylatticefer}) can be described by the following chiral transformation  on the original left-moving Dirac fermions of the continuous description (\ref{contlimitDirac}):
\begin{equation}
{\cal D}_{\rm dc}: \;  L_{1 \alpha}   \xrightarrow{{\cal D}_{\rm dc}}    i L_{1 \alpha} , \; \; 
 L_{2 \alpha}   \xrightarrow{{\cal D}_{\rm dc}}   - i L_{2 \alpha} ,
\label{densitycurrduality}
\end{equation}
the right-movers being invariant.  Using Eqs. (\ref{diagcurrentcont}, \ref{contOAFperp}) and (\ref{contcompetorderCDWBDW}), one can check that the duality (\ref{densitycurrduality}) gives the correct transformation of the order parameter (\ref{dualityCDWOAF}):
\begin{equation}
\begin{split}
 {\cal O}_{{\rm OAF}_{\perp}} &  \xrightarrow{{\cal D}_{\rm dc}}   - {\cal O}_{{\rm CDW}_{-}} , \\
 {\cal O}_{{\rm CDW}_{-}} &  \xrightarrow{{\cal D}_{\rm dc}}    {\cal O}_{{\rm OAF}_{\perp}} , \\
 {\cal O}_{{\rm OAF}_{\text d}} &  \xrightarrow{{\cal D}_{\rm dc}}   - {\cal O}_{{\rm BDW}_{-}} ,\\
 {\cal O}_{{\rm BDW}_{-}} &  \xrightarrow{{\cal D}_{\rm dc}}   {\cal O}_{{\rm OAF}_{\text d}}.
 \label{competorderduality}
\end{split}
\end{equation}

The density-current duality (\ref{densitycurrduality}) defined on the Dirac fermions leads to a $\mathbb{Z}_2$-grading of $\mathfrak{so}(4N)$ which reads as follows in terms of the left currents of the low-energy model (\ref{lowenergyhamhalfilling}):
\begin{equation}
\begin{split}
J^{A}_L &  \xrightarrow{{\cal D}_{\rm dc}}  J^{A}_L , J^{A,z}_L  \xrightarrow{{\cal D}_{\rm dc}}  J^{A,z}_L,  j_L^z   \xrightarrow{{\cal D}_{\rm dc}}  j_L^z, J_{cL}   \xrightarrow{{\cal D}_{\rm dc}}   J_{cL}, \\
J^{A,+}_L &  \xrightarrow{{\cal D}_{\rm dc}}  - J^{A,+}_L , j^{+}_L   \xrightarrow{{\cal D}_{\rm dc}}  - j^{+}_L , \\
S_L^{\alpha \beta +} &  \xrightarrow{{\cal D}_{\rm dc}}  S_L^{\alpha \beta +}, A_{ll L}^{\alpha \beta +}   \xrightarrow{{\cal D}_{\rm dc}}  - A_{ll L}^{\alpha \beta +}, 
A_{12 L}^{\alpha \beta +}   \xrightarrow{{\cal D}_{\rm dc}}  A_{12 L}^{\alpha \beta +} .
\end{split}
\label{currdualitydensitycurr}
\end{equation}
In Appendix \ref{dualityappen}, we show that this duality symmetry belongs to the symmetric 
class DIII of $\mathbb{Z}_2$-gradings of  $\mathfrak{so}(4N)$.  From the result (\ref{currdualitydensitycurr}), we deduce that the line $f_1 = - f_2 = f_3 = - f_4 = f_5 = f_6 = f_7 = - f_8 = f_9 = f >0$ of model (\ref{lowenergyhamhalfilling}) is 
related to the SO($4N$) GN model (\ref{GNso}) by the duality ${\cal D}_{\rm dc}$.  This line, dual to the SO($4N$)  GN model under ${\cal D}_{\rm dc}$, describes thus the formation of a long-range ordering of the  BDW$_{-}$  operator (\ref{BDW-para}) (see Fig. \ref{figcompetingorderslatt}).

The last case is the CDW$_{-}$ phase (\ref{CDW-}) which is dual to the OAF$_{\perp}$ phase by the density-current duality ${\cal D}_{\rm dc}$ (\ref{densitycurrduality}). It corresponds to the $f_1 = - f_2 = f_3 =  - f_4 = f_5 = f_6 = -f_7 = f_8 = -f_9 = f >0$ line by applying the ${\cal D}_{\rm dc}$ duality to the line  $f_1 = f_2 = f_3 =  f_4 = f_5 = f_6 = -f_7 = - f_8 = -f_9 = f >0$.  

\begin{figure}[t]
\begin{center}
\includegraphics[scale=0.8]{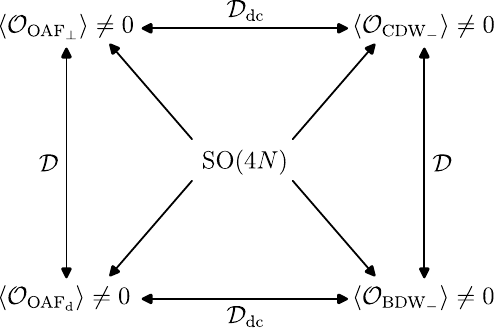}
\caption{(color online) The four competing orders related by two duality symmetries and unified by
the SO($4N$) symmetry of the Gross-Neveu model.}
\label{figcompetingorders}
\end{center}
\end{figure}

In summary, the four competing orders related by the lattice density-current duality  (\ref{dualitylatticefer})  and the duality 
${\cal D}$ (\ref{OAFduality}) are unified by the emergent SO($4N$) symmetry of the GN model (\ref{GNso}). 
They correspond to four different fully gapped-Mott insulating phases C0S0 with the long-range ordering of their order parameters and a two-fold ground-state degeneracy.  These phases are defined by the following line 
of parameters of the low-energy theory (\ref{lowenergyhamhalfilling}):
\begin{equation}
\begin{split}
\langle {\cal O}_{{\rm OAF}_{\perp}} \rangle \ne 0: 
& f_1 = f_2 = f_3 =  f_4 = f_5 = \\
& f_6 = -f_7 = - f_8 = -f_9 = f >0 , \\
\langle  {\cal O}_{{\rm CDW}_{-}}\rangle \ne 0:
& f_1 = - f_2 = f_3 =  - f_4 = f_5 = \\
& f_6 = -f_7 = f_8 = -f_9 = f >0, \\
\langle {\cal O}_{{\rm OAF}_{\text d}}  \rangle \ne 0:
& f_1 = f_2 = f_3 = f_4 = f_5 = \\
&  f_6 = f_7 =  f_8 = f_9 = f >0, \\
\langle  {\cal O}_{{\rm BDW}_{-}} \rangle \ne 0:
& f_1 = - f_2 = f_3 = - f_4 = f_5 = \\
& f_6 = f_7 = - f_8 = f_9 = f >0 .
\end{split}
\label{RGcompetingphaseshalfilling}
\end{equation}

Fig. \ref{figcompetingorders} summarizes the correspondence between these competing long-range ordered
phases, unified by the SO($4N$) symmetry of the GN model (\ref{GNso}), and related by 
the lattice density-current ${\cal D}_{\rm dc}$ duality symmetry and the emergent one ${\cal D}$, valid in the low-energy regime. Table \ref{tabsum} gives the discrete symmetries associated to the order parameters and the effect of the different duality symmetries.

\subsection{One-loop RG approach and phase diagrams}
\label{sec:1loopphasediaghalfilled}

The duality approach of Sec. \ref{sec:dualityGN} cannot predict whether the phases of Fig. \ref{figcompetingorders} 
occur in the phase diagram of the lattice model (\ref{eqn:hamlatticeladder}). In this respect, we need to solve numerically the one-loop RG equations (\ref{RGhalf}) with initial conditions (\ref{couplingshalfilling}) to observe if the RG flow is attracted along the special lines (\ref{RGcompetingphaseshalfilling}) in the far IR regime.

\subsubsection{Phase diagrams}
\label{sec:phasediaghalfilled}

The full analysis of the one-loop RG approach and 
the different phases of model (\ref{eqn:hamlatticeladder}) will be presented in a separate paper \cite{Habchy26}. Here, 
we investigate special surfaces of the lattice model (\ref{generalint}) where the phases (\ref{RGcompetingphaseshalfilling}) may be stabilized. We consider only the $N>2$ case, the $N=2$ case being very specific since the one-loop RG equations (\ref{RGhalf}) contain several $N-2$ factors. The RG analysis for $N=2$ at half-filling has already been discussed for instance in Refs. \onlinecite{Tsuchiizu-F-02,Nonne-B-C-L-10,Okamoto-M-12}.

\begin{figure}[t]
\begin{center}
\includegraphics[scale=0.9]{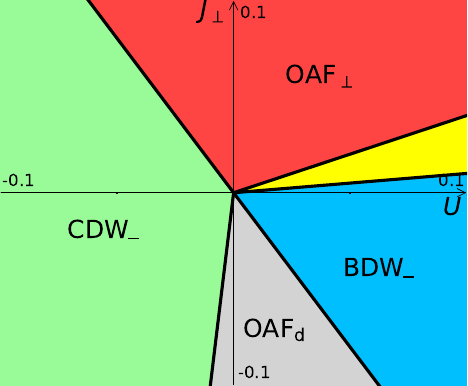}
\caption{(color online) RG phase diagram  for $V=V_{\rm pair}=0$ ($t=1$) and $N>2$ ($N=4$ here).}
\label{fighalfilling}
\end{center}
\end{figure}

Fig. \ref{fighalfilling} presents the zero-temperature phase diagram of model (\ref{doublewellint}) with $V=V_{\rm pair}=0$ at half-filling in the weak-coupling regime for $N>2$.
Interestingly enough, the four long-range ordered phases (\ref{RGcompetingphaseshalfilling}) are stabilized in 
this model.   The latter result signals the emergence of an enlarged SO($4N$) symmetry  in the far IR limit
along the four special lines $f_i = \epsilon_i f, \epsilon_i = \pm 1$ (\ref{RGcompetingphaseshalfilling}). It
is a rather well-known feature of 1D interacting Dirac fermions that very often the Hamiltonian with marginal relevant interactions is attracted under the RG flow to a manifold possessing a symmetry higher than that of the original one.  \cite{Lin-B-F-98} This phenomenon is called dynamical symmetry enlargement (DSE) and many examples have been discussed before within the one-loop RG approach \cite{Lin-B-F-98,Konik-S-L-02,Assaraf-A-B-C-L-04,Boulat-A-L-09,Lee-A-B-04,Lecheminant-B-A-05,Lecheminant-T-06-SU4,Lecheminant-T-06-SDSG,Lecheminant-A-B-08}.
In stark contrast, the unlabeled yellow phase in Fig. \ref{fighalfilling} corresponds to the  RG ray: 
$ f_1 = - f_3 = f_6 = - f_7 = f_9 =  f >0$ with  $f_{2,4,5,8}/f \rightarrow 0$, which is not described by a SO($4N$) DSE phenomenon.  This  phase is a C1S0 phase with gapless relative-charge degrees of freedom.
It corresponds to the formation of a BDW$_{\perp}$ phase which is characterized by algebraic correlation function
of its order parameter (\ref{CDWODDxlattice}) \cite{Habchy26}.

\begin{figure}[t]
\begin{center}
\includegraphics[scale=0.9]{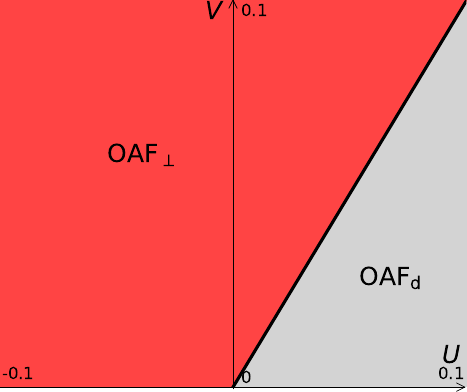}
\caption{(color online) RG phase diagram  for model ${\rm H}^{{\cal D}_{\rm dc}}_{\rm dw}$ (\ref{dualdw}) with
$N>2$ ($t=1$ and $N=4$ here).}
\label{figdualcoldatoms}
\end{center}
\end{figure}

In the weak-coupling regime, we also find that the half-filled SU($N$) Hubbard two-leg ladder (\ref{hubbard2leg})  exhibits two C0S0 phases depending on the sign of $U$: a ${\rm BDW}_{-}$ (respectively ${\rm CDW}_{-}$) phase for repulsive (respectively attractive) $U$ (see the $J_{\perp} =0$ axis of Fig. \ref{fighalfilling}).
Using the lattice density-current duality symmetry ${\cal D}_{\rm dc}$(\ref{dualitylatticefer}), we deduce  that its dual-lattice model (\ref{dualHubbard}) harbors two loop-current ordered phases,  an ${\rm OAF}_{\perp}$ phase and ${\rm OAF}_{\text d}$ phase for respectively $U<0$ and $U>0$. Fig. \ref{figdualcoldatoms} presents the RG analysis for 
 model (\ref{dualdw}) with $N>2$ which includes the dual Hubbard two-leg ladder  (\ref{dualHubbard}) when $V=0$.  For repulsive $V > 0$, the two loop-current ordered phases do appear and they are separated by a quantum phase transition located at $U=V$.

The existence of a staggered bond-ordered phase ${\rm BDW}_{-}$ for the half-filled SU($N$) Hubbard two-leg ladder  is expected to be a property of the weak-coupling regime and should not survive in the large-$U$ limit. Indeed, using the strong-coupling approach of Ref. \onlinecite{Fromholz-C-L-P-T-19}, it is known that a non-degenerate fully-gapped rung singlet phase emerges in the large-$U$ limit. A quantum phase transition as function of $U$ should then take place between the ${\rm BDW}_{-}$ and singlet phases. Using the duality symmetry ${\cal D}_{\rm dc}$, we expect that the ${\rm OAF}_{\text d}$ phase is only present in the phase diagram of the dual model (\ref{dualHubbard}) in a finite region of $U>0$. In stark contrast, the ${\rm CDW}_{-}$ phase persists in the large attractive $U$ regime of  the half-filled SU($N$) Hubbard two-leg ladder \cite{Fromholz-C-L-P-T-19}.  The current-ordered ${\rm OAF}_{\perp}$ phase  is thus formed for all negative values of $U$ in  model (\ref{dualHubbard}).

\subsubsection{Self-dual manifolds}
\label{sec:selfdualhalfilled}

The duality symmetries found above enable us to discuss the nature of the quantum phase transitions between
the different phases by investigating self-dual manifolds.  

We start with the self-dual manifold related to the first duality ${\cal D}$ (\ref{OAFduality}). This duality is emergent at low-energy and does not exist on the lattice in stark contrast to the density-current duality symmetry ${\cal D}_{\rm dc}$. 
The self-dual manifold associated to ${\cal D}$ in the low-energy approach corresponds to $g_7 = g_8 =g_9=0$ according to Eq. (\ref{currdualitytildeOmega}). The low-energy Hamiltonian density which governs the ${\cal D}$ self-dual manifold
with $f_1 = f_2 =  f_3 = f_4  = f_5 = f_6 = f>0$ is given by:
\begin{equation}
\begin{split}
{\cal H}^{\Omega}_{{\cal D} \; \rm{self-dual}} & = -i v_\text{F}\left(:R_{l \alpha} ^{\dagger} \partial_x R_{l \alpha}:  - 
  :L_{l \alpha}^{\dagger}  \partial_x L_{l \alpha}:  \right)  \\
  & + \frac{\pi f}{2 N^2}  J_{cR} J_{cL}   + \frac{2\pi f}{N}  \mathcal{J}_{L}^{A}   \mathcal{J}_{R}^{A} ,
\end{split}
\label{HselfdualOmegatilde}
\end{equation}
$\mathcal{J}_{L,R}^{A}$ ($A= 1, \ldots, 4N^2-1$) being the chiral SU($2N$)$_1$ currents introduced in
Eqs. (\ref{SU(2N)_1curr}).

This Hamiltonian density displays a ``spin-charge''  separation between the U(1)$_{\rm c}$ uniform charge degrees of freedom and the remaining SU($2N$) degrees of freedom:
\begin{equation}
\begin{split}
 {\cal H}^{\Omega}_{{\cal D}\;  \rm{self-dual}}  & = {\cal H}_c 
+ {\cal H}_s, \\
 \left[  {\cal H}_{\rm c}, {\cal H}_{\rm s} \right] &= 0, 
\end{split}
\label{HselfdualOmegatildeLuttingerGNbis}
\end{equation}
with
\begin{equation}
\begin{split}
{\cal H}_c  &= \frac{\pi v_{F}}{2N} \left( : J_{cL}^2: +  : J_{cR}^2: \right) + \frac{\pi f}{2 N^2}  J_{cR} J_{cL} ,   \\
{\cal H}_s & =  \frac{2\pi v_{F}}{2N+1} \left( : \mathcal{J}_{L}^{A} \mathcal{J}_{L}^{A}: + 
: \mathcal{J}_{R}^{A} \mathcal{J}_{R}^{A}:  \right) + \frac{2\pi f}{N}  \mathcal{J}_{L}^{A}   \mathcal{J}_{R}^{A} ,
\end{split}
\label{HselfdualOmegatildeLuttingerGN}
\end{equation}
where the Sugawara decomposition (\ref{HamcontDiracurrentsU(2N)}) has been used. In Eq. (\ref{HselfdualOmegatildeLuttingerGN}), ${\cal H}_c$ corresponds to the Tomonaga-Luttinger liquid Hamiltonian
density with gapless $c=1$ behavior which describes 1D metallic systems \cite{Giamarchi,Gogolin-N-T-book}. 
The spin part, ${\cal H}_s$, in Eq. (\ref{HselfdualOmegatildeLuttingerGN}) corresponds to the SU($2N$)  chiral GN or non-Abelian Thirring model \cite{Gross-N-74}. The latter model is a massive integrable field theory
when $f>0$ \cite{Andrei-L-80}.  We thus expect that the self-dual manifold is  a $c=1$ CFT governed by the uniform charge degrees of freedom. 

 The second  self-dual manifold is related to the density-current duality ${\cal D}_{\rm dc}$. 
The self-dual manifold on the lattice is defined by Eq. (\ref{selfdualine}) or by $g_2 = g_4 =g_8=0$ within the low-energy approach. The one-loop RG equations along the self-dual manifold are then given by Eq. (\ref{RGhalf}) with 
$g_2 = g_4 =g_8=0$.  In particular, one finds that $g_5$ remains constant. 
The quantum phase transition between the ${\rm OAF}_{\perp}$ (respectively ${\rm OAF}_{\text d}$  ) and ${\rm CDW}_{-}$ (respectively ${\rm BDW}_{-}$) phases belong to this 
self-dual manifold with $f_1 = f_3 = f_6 = -f_7  = -f_9 = f >0$. The low-energy Hamiltonian density which controls the quantum phase transition between these phases is given by:
\begin{equation}
\begin{split}
& {\cal H}^{\Omega}_{{\cal D}_{\rm dc} \; \rm{self-dual}}  = -i v_{F}\left(:R_{l \alpha} ^{\dagger} \partial_x R_{l \alpha}:  - :L_{l \alpha}^{\dagger}  \partial_x L_{l \alpha}:  \right) \\
  & +  \frac{2\pi f}{N}  \left( J_{1L}^A J_{1R}^A + J_{2L}^A J_{2R}^A \right) + \frac{\pi f}{2 N^2}  J_{cR} J_{cL}   
  + \frac{2\pi f_5}{N^2}  j_L^z  j_R^z  \\
  & - \frac{\pi f}{N}  \left( L^{\dagger}_{1 \alpha} R_{1 \alpha}  L^{\dagger}_{2 \beta}  R_{2 \beta}  +  {\rm H.c.} \right) .
\end{split}
\label{HselfdualOmegadc}
\end{equation}
As in Eq. (\ref{HselfdualOmegatildeLuttingerGNbis}), this Hamiltonian can be separated into two commuting pieces:
\begin{equation}
\begin{split}
 {\cal H}^{\Omega}_{{\cal D}_{\rm dc} \; \rm{self-dual}}  &= {\cal H}_{\rm c-} + {\cal H}_{\rm sc} ,\\
\left[  {\cal H}_{\rm c-}, {\cal H}_{\rm sc} \right] & = 0 , 
\end{split}
\label{HselfdualOmegadcLuttingerbis}
\end{equation}
with
\begin{equation}
\begin{split}
& {\cal H}_{\rm c-}  = \frac{2 \pi v_{F}}{N} \left( : (j_L^z)^2: +: (j_R^z)^2:  \right) + \frac{2\pi f_5}{N^2}  j_L^z  j_R^z  ,   \\
& {\cal H}_{\rm sc}  =   \frac{\pi v_{F}}{2 N} \left( : J_{cL}^2: + : J_{cR}^2: \right) +
\frac{2\pi v_{F}}{N+1}  \left(  :J_{1L}^A J_{1L}^A: +  \right. \\
& \left. + :J_{2 L}^A J_{2 L}^A: + :J_{1R}^A J_{1R}^A: +  :J_{2 R}^A J_{2 R}^A: \right)   \\
& + \frac{\pi f}{2 N^2}  J_{cR} J_{cL} +  \frac{2\pi f}{N}  \left( J_{1L}^A J_{1R}^A + J_{2L}^A J_{2R}^A \right)  \\
& - \frac{\pi f}{N}  \left( L^{\dagger}_{1 \alpha} R_{1 \alpha}  L^{\dagger}_{2 \beta}  R_{2 \beta}  +  {\rm H.c.} \right) .
\end{split}
\label{HselfdualOmegadcLuttinger}
\end{equation}
The first piece, i.e. ${\cal H}_{\rm c-}$, takes the form of a  Tomonaga-Luttinger liquid Hamiltonian
density for the orbital or relative charge degrees of freedom related to the U(1)$_{\rm o}$ symmetry. A gapless $c=1$ behavior emerges from this contribution. In contrast, ${\cal H}_{\rm sc}$, which describes the coupling between 
charge and SU($N$) spin degrees of freedom, has a massive low-energy spectrum.  Indeed, as seen in Eq.
(\ref{HselfdualOmegadcLuttinger}), each SU($N$) degree of freedom in channel $l=1,2$ is described by a 
massive integrable field theory which is the SU($N$) chiral GN model with coupling constant $f>0$. A spin gap  is thus formed for the spin degrees of freedom.  In the low-energy regime below this spin gap, averaging over the spin degrees of freedom,  one finds that ${\cal H}_{\rm sc}$ takes the form of a sine-Gordon model for the remaining uniform charge degrees of freedom:
\begin{equation}
{\cal H}_{\rm sc} \simeq \frac{v_c}{2} \left[\frac{1}{K_c} 
\left(\partial_x \Phi_c \right)^2 + K_c 
\left(\partial_x \Theta_c \right)^2  \right] - g_c \cos \left( \sqrt{\frac{8 \pi}{N}} \Phi_c \right),
\label{HselfdualSGcharge}
\end{equation}
where  the uniform charge excitations are described by the bosonic field $\Phi_c$ and its dual field $\Theta_c$, and 
$v_c, K_c$  are the Luttinger parameters.  The sine-Gordon model (\ref{HselfdualSGcharge}) has a charge gap when $K_c < N$, which is the case here. The low-energy Hamiltonian density (\ref{HselfdualOmegadc}), which describes
the self-dual manifold related to the density-current duality ${\cal D}_{\rm dc}$, has thus a $c=1$ gapless behavior which stems from the U(1)$_{\rm o}$ symmetry.  The action of the latter symmetry on the original Dirac fermions of 
the continuum limit of the band lattice fermions (\ref{contlimitDirac}) is given by:
\begin{equation}
\begin{split}
& L_{1 \alpha}  \xrightarrow{{\rm U(1)}_{\rm o}}  e^{ i \theta/2} L_{1 \alpha}, \; 
R_{1 \alpha}  \xrightarrow{{\rm U(1)}_{\rm o}}  e^{ i \theta/2} R_{1 \alpha}, \\
& L_{2 \alpha}  \xrightarrow{{\rm U(1)}_{\rm o}}  e^{ - i \theta/2} L_{2 \alpha}, \; 
R_{2 \alpha}  \xrightarrow{{\rm U(1)}_{\rm o}}  e^{ - i \theta/2} R_{2 \alpha} .
 \end{split}
 \label{U1osym}
\end{equation}
Using the expressions of the order parameters (\ref{diagcurrentcont}), (\ref{contOAFperp}) and (\ref{contcompetorderCDWBDW}) in terms of the Dirac fermions, one can show that the 
 U(1)$_{\rm o}$ symmetry rotates each pair of dual order parameters as in Eqs. (\ref{doubletorderpara}) 
 and (\ref{doubletorderpara2}). The dual orders are continuously transformed to each other and disappear at the transition with the emergence of the U(1)$_{\rm o}$ symmetry which is gapless.

The quantum phase transition between the ${\rm OAF}_{\perp}$ and 
${\rm OAF}_{\text d}$ phases in Fig. \ref{figdualcoldatoms} has the particularity to occur along the $U=V>0$ line of model (\ref{dualdw}). The initial conditions of the RG approach (\ref{couplingshalfilling}) give $g_2 = g_4= g_7= g_8 = g_9 =0$. The $U=V$ line is thus invariant under both ${\cal D}$ and ${\cal D}_{\rm dc}$ duality symmetries. The RG equations (\ref{RGhalf}) can be exactly solved along this self-dual manifold and  one finds $g_{1,3} \rightarrow 0$ in the far IR limit.  We deduce that all degrees of freedom are gapless along this transition line. The quantum phase transition is  continuous with central charge $c=2N$ in 
the weak-coupling regime.

\section{Doped case}
\label{sec:low-energydoped}
We now consider the lightly-doped case by considering a generic filling close to
half-filling. We investigate the doping of the long-range ordered quantum phases (\ref{RGcompetingphaseshalfilling})
by means of a one-loop RG approach.

\subsection{RG approach for lightly-doped ladder}
\label{sec:RGdoped}
For a generic incommensurate filling $f$ away from half-filling, one has  $k_{1F} + k_{2F} = 2\pi f /a_0$ and the anisotropy $\delta = a_0 (k_{2F} - k_{1F}) $ satisfies $\sin( \delta/2) = t_{\perp}/(2t \sin(\pi f)) $ with  $t_{\perp} < 2 t \sin(\pi f)$ 
to get four Fermi points.  Away from half-filling, there is now two different Fermi velocities. In the lightly-doped case, by considering a generic filling close to half-filling, the velocity anisotropy of the Fermi velocities can be  safely neglected. 

At generic fillings, the two Fermi momenta are incommensurate and no additional operators to treat umklapp processes at  special dopings are required as in the half-filled case. For lightly-doped ladder, the interacting Hamiltonian density of the low-energy approach is then given by Eq. (\ref{lowenergyhamhalfilling}) with $g_7 = g_8 = g_9 =0$:
\begin{equation}
\begin{split}
 \mathcal{H}^{\Omega}_{\text{int}} 
&= g_1 J_\text{L}^A J_\text{R}^A +g_2 \left(J_{L}^{A,x}  J_{R}^{A,x} + J_{L}^{A,y}  J_{R}^{A,y} \right) \\
& + g_3 J_\text{L}^{A,z} J_\text{R}^{A,z}  
   + g_4 \left(j_{L}^x  j_{R}^x  + j_{L}^y  j_{R}^y \right)  \\
 & +g_5 j_{\text{L}}^z j_{\text{R}}^z  + g_6  J_{cL} J_{cR}  ,
\end{split}
  \label{lowenergyhamchiraltrans}
\end{equation}
with the initial conditions of Eq. (\ref{couplingshalfilling}). 

Model (\ref{lowenergyhamchiraltrans}) enjoys a ``spin-charge '' separation since the U(1)$_c$ charge currents 
$J_{cL,R}$ commute with the remaining four-fermion terms which describe fluctuations in the spin and orbital sectors. 
The charge degrees of freedom, described by the last term in Eq. (\ref{lowenergyhamchiraltrans}), decouples from the rest of the interaction. For incommensurate filling, the uniform charge degrees of freedom display metallic properties in the Luttinger liquid universality class which are governed by the Tomonaga-Luttinger  density Hamiltonian:\cite{Gogolin-N-T-book,Giamarchi} 
\begin{equation}
{\cal H}_c = \frac{v_c}{2} \left[\frac{1}{K_c} 
\left(\partial_x \Phi_c \right)^2 + K_c 
\left(\partial_x \Theta_c \right)^2 \right].
\label{luttbis}
\end{equation}
The explicit form of the Luttinger parameters in the weak-coupling regime can be extracted from the continuum limit of
the model:
\begin{eqnarray}
K_c &=&  \frac{1}{\sqrt{1 + 2 N g_6/\pi v_F}} ,
\nonumber \\
v_c &=& v_F\sqrt{1 + 2 N g_6/\pi v_F} ,
\label{Luttingerparameters}
\end{eqnarray}
with $g_6 = a_0 ( \frac{N-1}{2N} U + \frac{V}{2})$. 

The remaining spin and orbital (relative-charge) degrees of freedom interact through the density Hamiltonian (\ref{lowenergyhamchiraltrans}) whose one-loop RG equations are given by 
Eq. (\ref{RGhalf}) with $g_7=g_8 = g_9 =0$:
\begin{equation}
\begin{split}
\dot{g_1} =& \frac{N}{4\pi} g^2_1 + \frac{N}{8\pi} g^2_2 + \frac{N}{16\pi} g^2_3,  \\
\dot{g_2} =&  \frac{N}{2\pi} g_1 g_2 + 
    \frac{N^2 -4}{4\pi N} g_2 g_3 +   \frac{1}{2\pi} (g_2 g_5+g_3 g_4) ,  \\
\dot{g_3} =&  \frac{N}{2\pi} g_1 g_3 +  \frac{N^2 -4}{4\pi N} g^2_2 +    \frac{1}{\pi} g_2 g_4,  \\
\dot{g_4} =&  \frac{1}{2\pi} g_4 g_5 + \frac{N^2 - 1}{2\pi N^2} g_2 g_3 ,  \\
 \dot{g_5} =& \frac{N^2 -1}{2\pi N^2} g^2_2 + \frac{1}{2\pi} g^2_4   .
\end{split}
\label{RGinc}
\end{equation}

Solving numerically these RG equations for $N>2$, we find the following RG rays which include two highly-symmetric rays obtained in the  half-filled case (\ref{RGcompetingphaseshalfilling}) with $f_7 = f_8 = f_9 =0$:
 \begin{equation}
\begin{split}
&{\rm I}: \; \; f_1 = f_2 = f_3 = f_4 = f_5 =  f > 0 , \\
& {\rm II}: \; \; f_1 = - f_2 = f_3 = -f_4 = f_5 =  f > 0 , \\
& {\rm III}: \; \; f_1 = - f_3 =  f > 0, \; \frac{f_{2,4,5}}{f} \rightarrow 0 ,\\
& {\rm IV}: \; \;  f_5 = - f < 0, \frac{f_{1,2,3, 4}}{f} \rightarrow 0 . \\
\end{split}
\label{RGflow}
\end{equation}

\begin{figure}[t]
\begin{center}
\includegraphics[scale=1]{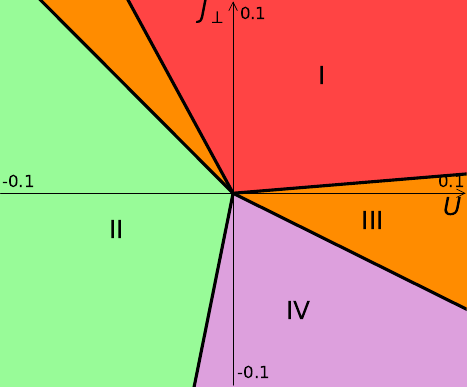}
\caption{(color online) RG phase diagram  for $V=V_{\rm pair}=0$ ($t=1$) for incommensurate fillings and $N>2$
($N=4$ here).}
\label{figinc}
\end{center}
\end{figure}

Fig. \ref{figinc} presents the weak-coupling phase diagram of the two-leg SU($N$) ladder (\ref{generalint}) with $V=V_{\rm pair}=0$ and $N>2$ for incommensurate fillings close to half-filling. Phase IV describes a generalized Luttinger phase where all modes are gapless. In the region III of RG rays (\ref{RGflow}), there is a spin gap and the two charge modes remain critical. It signals the formation of a C2S0 phase.
The physical properties of this phase, as well as a more careful treatment of the velocity-anisotropy, will be addressed in 
a separate publication~\cite{Habchy26}.

\subsection{Loop-current phases}
\label{sec:Phase I}

We focus on phase I in Fig. \ref{figinc} which should govern the doping effects of the loop-current ordered phase 
OAF$_{\perp}$ found in the half-filled case. Along the line $f_i = f >0$, model~(\ref{lowenergyhamchiraltrans}) enjoys a DSE phenomenon which is characterized by an enlarged SU($2N$) symmetry in the far IR. The physical properties of the phase are captured by the SU($2N$)  chiral GN or non-Abelian Thirring model \cite{Gross-N-74}, already introduced in Eq. (\ref{HselfdualOmegatildeLuttingerGN}):
\begin{equation}
\begin{split}
{\cal H}_{\rm{chiral} \; \rm{GN}} & = -i v_\text{F}\left(:R_{l \alpha} ^{\dagger} \partial_x R_{l \alpha}:  - 
  :L_{l \alpha}^{\dagger}  \partial_x L_{l \alpha}: \right)  \\
  &+ \frac{2\pi f}{N}  \mathcal{J}_{L}^{A}   \mathcal{J}_{R}^{A} .
 \end{split}
\label{GN}
\end{equation}

The chiral  GN model (\ref{GN}) is a massive integrable field theory when $f>0$ in the SU($2N$) spin sector  \cite{Andrei-L-80}. We have a thus a non-perturbative generation of a gap and the phase I is gapless with central charge $c=1$.
It signals the formation of a C1S0 phase governed by the charge degrees of freedom (\ref{luttbis})  that are decoupled from the remaining SU($2N$) degrees of freedom.   

The nature of the physical properties of this phase can be obtained by means of the Abelian bosonization approach 
which is described in Appendix  \ref{bosonisationappen}.  After performing the chiral transformation $\Omega$ (\ref{chiraltrans}) on  the IR Hamiltonian (\ref{GN}) along the isotropic line $f_i =f$, the interacting Hamiltonian can 
be bosonized using the identification (\ref{bosoabeleq}) and the correspondence (\ref{finalident}):
\begin{equation}
\begin{split}
&  {\cal H}_{\rm int}^{(I)} =  - \frac{f}{N} \left( \partial_x \Phi_{oL} \partial_x \Phi_{oR} 
- \partial_x {\vec \Phi}_{+L} \cdot  \partial_x {\vec \Phi}_{+R}  \right. \\
& \left. +  \partial_x {\vec \Phi}_{-L} \cdot \partial_x  {\vec \Phi}_{-R} \right)  \\
& - \frac{f}{2 \pi N a_0^2} \sum_{\alpha}  
\cos \left( \sqrt{\frac{8 \pi}{N}} \Theta_o + \sqrt{8 \pi} {\vec e}_{\alpha} \cdot {\vec \Theta}_{-}  \right) \\
& + \frac{f}{2 \pi N a_0^2} \sum_{\alpha \ne \beta}  \Gamma_{\alpha \beta} 
\cos \left( \sqrt{2 \pi} \left( {\vec e}_{\alpha} -  {\vec e}_{\beta} \right) \cdot {\vec \Phi}_{+}  \right) \\
& \times \left[ \cos \left( \sqrt{2 \pi} \left( {\vec e}_{\alpha} -  {\vec e}_{\beta} \right)  \cdot {\vec \Theta}_{-} \right) \right. \\
& \left.
+ \cos \left( \sqrt{\frac{8 \pi}{N}} \Theta_o + \sqrt{2 \pi} \left( {\vec e}_{\alpha} +  {\vec e}_{\beta} \right)
\cdot {\vec \Theta}_{-} \right) \right],
\end{split}
\label{bosonizedHregionI}
\end{equation}
where the charge part, that decouples from the interaction, is not included and we have introduced
$\Gamma_{\alpha \beta} = \kappa_{1 \alpha}\kappa_{2 \alpha}\kappa_{1 \beta}\kappa_{2 \beta}$. This operator which depends on the Klein factors (see Appendix \ref{bosonisationappen}) satisfies: $\Gamma_{\alpha \beta}^2 = 1$ and commutes with $ {\cal H}_{\rm int}^{(I)} $.
In this respect, we can then fix this product of these Klein factors such that $\Gamma_{\alpha \beta} = -1$ for instance.  Since the Hamiltonian (\ref{GN}), which describes the IR physics of phase I, has a non-perturbative spectral gap
for all degrees of freedom except the uniform charge ones,  all bosonic fields of Eq. (\ref{bosonizedHregionI}) are pinned in the ground state.   We can choose, for instance, in the ground 
state the pinning configurations for the bosons: $\langle \Theta_o \rangle = 0, \langle {\vec \Phi}_{+}  \rangle = {\vec 0}$, and $\langle {\vec \Theta}_{-}  \rangle = {\vec 0}$.
 
Using the continuum description (\ref{contlimitDirac}) of the lattice fermions in terms of the Dirac ones, one can derive the continuum limit of the lattice order parameter  (\ref{ODWy}): 
\begin{equation}
 \begin{split}
 {\cal O}_{{\rm OAF}_{\perp}} & \simeq  \frac{i}{2} \left( e^{i 2k_F x} L_{2\alpha}^\dagger R_{1 \alpha} 
 + e^{-i 2k_F x} R_{2\alpha}^\dagger L_{1 \alpha}  \right. \\
 & \left. - e^{i 2k_F x} L_{1\alpha}^\dagger R_{2 \alpha} 
 - e^{-i 2k_F x} R_{1\alpha}^\dagger L_{2 \alpha}   \right),
\end{split}
\label{ODWycont}
\end{equation}
where $k_F = (k_{1F} + k_{2 F})/2$ and the chiral terms have been ignored since they will give exponential decaying contributions to the two-point function of the order parameter.
The next step of the approach is to express the order parameter (\ref{ODWycont}) in terms of the bosonic fields which 
occur in Eq. (\ref{bosonizedHregionI}). We find: 
\begin{equation}
 \begin{split}
& {\cal O}_{{\rm OAF}_{\perp}}  \simeq  \frac{1}{\pi a_0} \cos \left( 2k_F x + \sqrt{\frac{2 \pi}{N}} \Phi_c \right)
 \sum_{\alpha} i \kappa_{2 \alpha} \kappa_{1 \alpha}   \\
& \times\cos \left( \sqrt{2 \pi} {\vec e}_{\alpha} \cdot
 {\vec \Phi}_{+}\right)  \cos \left( \sqrt{\frac{2 \pi}{N}} \Theta_{o} + \sqrt{2 \pi} {\vec e}_{\alpha} \cdot
 {\vec \Theta}_{-}\right)  \\
 & - \frac{1}{\pi a_0} \sin \left( 2k_F x + \sqrt{\frac{2 \pi}{N}} \Phi_c \right)
 \sum_{\alpha} i \kappa_{2 \alpha} \kappa_{1 \alpha}  \\
&  \times\sin \left( \sqrt{2 \pi} {\vec e}_{\alpha} \cdot
 {\vec \Phi}_{+}\right)  \cos \left( \sqrt{\frac{2 \pi}{N}} \Theta_{o} + \sqrt{2 \pi} {\vec e}_{\alpha} \cdot
 {\vec \Theta}_{-}\right)  .
\end{split}
\label{ODWyboson}
\end{equation}
In the low-energy regime $E \ll \Delta$,  $\Delta$ being the gap of the chiral GN model (\ref{GN}), one can average over the fast fields, the bosons $\Theta_o, {\vec \Phi}_{+}$ and $ {\vec \Theta}_{-} $, which are pinned in the ground-state  of the interacting Hamiltonian density (\ref{bosonizedHregionI}) with 
$\langle \Theta_o \rangle = 0, \langle {\vec \Phi}_{+}  \rangle =  \langle {\vec \Theta}_{-}  \rangle = {\vec 0}$.
The low-energy description of the order parameter (\ref{ODWyboson}) depends then  only on the massless charge bosonic field $\Phi_c$:
\begin{equation}
{\cal O}_{{\rm OAF}_{\perp}}  \simeq  \frac{C}{\pi a_0}  \sum_{\alpha} i \kappa_{2 \alpha} \kappa_{1 \alpha}  \cos \left( 2k_F x + \sqrt{\frac{2 \pi}{N}} \Phi_c \right),
\label{ODWybosonfin}
\end{equation}
where $C$ is a non-universal constant related to the different vacuum expectation values of the pinned bosons
in Eq. (\ref{ODWyboson}). We deduce the equal-time two-point function of the order parameter:
\begin{equation}
\langle {\cal O}_{{\rm OAF}_{\perp}} (x)   {\cal O}_{{\rm OAF}_{\perp}} (0) \rangle \simeq  \frac{N^2 C^2}{2 \pi^2 a_0^2}  \frac{ \cos ( 2k_F x)}{x^{\frac{K_c}{N}}},
\label{ODWycorrel}
\end{equation}
which has a non-universal algebraic decay which depends on the interaction from 
its Luttinger parameter $K_c$ (\ref{Luttingerparameters}).

In close parallel to the analysis of the transverse-current ordered parameter (\ref{ODWyboson}), one can derive the low-energy  description of the in-chain currents (\ref{OAFpara}) within the bosonization approach. 
One finds after exploiting the pinning of the  bosons $\Theta_o, {\vec \Phi}_{+}$ and $ {\vec \Theta}_{-} $:
\begin{equation}
\begin{split}
{\cal O}_{{\rm OAF}_{\parallel}}   \simeq  \frac{C}{4 \pi a_0} &  \sum_{\alpha} i \kappa_{1 \alpha} \kappa_{2 \alpha} 
\left[ e^{i 2k_F x + i \sqrt{\frac{2\pi}{N}} \Phi_c}  \left(e^{i k_{2F} a_0}   \right.  \right. \\
& \left. \left. - e^{i k_{1F} a_0}\right)   + {\rm H.c.} \right],
\end{split}
\label{OAFparabosonfin}
\end{equation}
with equal-time two-point function:
\begin{equation}
\begin{split}
\langle {\cal O}_{{\rm OAF}_{\parallel}} (x)   {\cal O}_{{\rm OAF}_{\parallel}} (0) \rangle & \simeq  
 \frac{ N^2 C^2}{2\pi^2 a_0^2}  \sin^2(\frac{\delta}{2}) \frac{ \cos ( 2k_F x)}{x^{\frac{K_c}{N}}}  \\
& \simeq   \frac{ N^2 C^2}{8\pi^2 a_0^2}  \frac{t_{\perp}^2}{t^2 \sin^2(\pi f)}   \frac{ \cos ( 2k_F x)}{x^{\frac{K_c}{N}}} ,
\end{split}
\label{OAFparacorrel}
\end{equation}
which has a similar leading asymptotics as in Eq. (\ref{ODWycorrel}) but with a different amplitude.

We now consider the OAF$_{\rm d}$ order parameter (\ref{diagcurrent}) corresponding to the circulation of the currents along the diagonals of the ladder.  Its continuum description reads as follows:
\begin{equation}
 \begin{split}
 {\cal O}_{{\rm OAF}_{\rm d}} & \simeq  - \frac{i}{4} \left( e^{i 2k_F x + i k_{1F} a_0} L_{1\alpha}^\dagger R_{2 \alpha}
  \right. \\
 & \left. + e^{- i 2k_F x - i k_{1F} a_0} R_{1\alpha}^\dagger L_{2 \alpha}  \right. \\
 & \left.  - e^{i 2k_F x + i k_{2F} a_0} L_{2 \alpha}^\dagger R_{1 \alpha} \right. \\
  & \left.  - e^{-i 2k_F x - i k_{2F} a_0} R_{2\alpha}^\dagger L_{1 \alpha}  - {\rm H.c.}  \right) .
\end{split}
\label{OAFdcont}
\end{equation}
Using the bosonization approach and averaging over the bosons $\Theta_o, {\vec \Phi}_{+}$ and $ {\vec \Theta}_{-} $,  one finds:
\begin{equation}
\begin{split}
{\cal O}_{{\rm OAF}_{\rm d}}   \simeq  \frac{C}{4 \pi a_0} &  \sum_{\alpha} i \kappa_{2 \alpha} \kappa_{1 \alpha} 
\left[ e^{i 2k_F x + i \sqrt{\frac{2\pi}{N}} \Phi_c}  \left(e^{i k_{2F} a_0} \right.  \right. \\
& \left. \left. + e^{i k_{1F} a_0}\right)  
 + {\rm H.c.} \right] .
\end{split}
\label{OAFdbosonfin}
\end{equation}
From this expression, we deduce the equal-time correlation function of this order parameter:
\begin{equation}
\begin{split}
\langle {\cal O}_{{\rm OAF}_{\rm d}}(x)   {\cal O}_{{\rm OAF}_{\rm d}} (0) \rangle  \simeq  
 \frac{ N^2 C^2}{2\pi^2 a_0^2}  \cos^2(\frac{\delta}{2}) \frac{ \cos ( 2k_F x)}{x^{\frac{K_c}{N}}} .
\end{split}
\label{OAFdcorrel}
\end{equation}

In phase I, we have the coexistence of the OAF phases defined by 
the transverse-current ordered parameter (\ref{ODWy}) with currents circulating along the ladder and the one with diagonal currents (\ref{diagcurrent}). Both instabilities have the same leading asymptotics and the $2k_F$ modulation. They differ by 
the amplitude of their two-point correlation functions, the transverse one (\ref{ODWy}) being larger.
 In stark contrast,  order parameters as the  $2k_F$ uniform/relative CDW, bond-density waves and 
charge $2e$ superconducting instabilities are fluctuating orders with exponential decaying correlation functions in phase I. 

A $4k_F$ CDW instability may coexist with the OAF phases.  The bosonized description of the $4k_F$ CDW can be derived as in Refs. \onlinecite{Chang-A-07,Robinson-al-12} by taking into account the higher-order contributions in perturbation theory of the density operator $n(x)$. A much simpler approach to obtain the continuous description 
of the $4k_F$ component of the CDW operator  is to consider the order parameter $n^2$, following the proposal of Ref. \onlinecite{White-A-S-02} in the context of lightly doped two-leg t-J ladder. The latter order parameter contains terms like $L^{\dagger}_{1 \alpha} R_{1 \alpha} L^{\dagger}_{2 \beta} R_{2 \beta}$ with 
momentum $ 4 k_F = 2k_{1F} +  2k_{2 F}$.
Using the bosonization approach and averaging over the pinned bosons in phase I
$\langle \Theta_o \rangle = 0, \langle {\vec \Phi}_{+}  \rangle = \langle {\vec \Theta}_{-}  \rangle = {\vec 0}$,  we find that 
the $4k_F$ CDW operator has a power-law correlation function in phase I with the leading asymptotics:
\begin{equation}
\langle n^2 (x) n^2 (0) \rangle  \simeq  \frac{N^2 C^2}{2 \pi^4 a_0^4} \;  \frac{ \cos \left(4k_F x \right)}{x^{\frac{4 K_c}{N}}} .
\label{4kFCDWcorrel}
\end{equation}
The correlation is less dominant than the two-point function of the OAF$_{\perp}$ order parameter (\ref{ODWycorrel}). The $4k_F$ CDW coexist with the OAF phases but the leading instability in phase I is OAF$_{\perp}$. As seen in Fig. \ref{figinc}, this phase is stabilized for repulsive $U$ and antiferromagnetic spin-exchange interaction $J_{\perp} > 0$ with $V=V_{\rm pair} =0$ and $N>2$.

\subsection{CDW$_{-}$ and BDW$_-$ phases}
\label{sec:Phase II}
 
The second asymptotic line of the RG (\ref{RGflow}) describes a DSE with SU($2N$) symmetry in the far IR limit.
Indeed, by  considering the density-current duality ${\cal D}_{\rm dc}$ (\ref{densitycurrduality}) symmetry,
the Hamiltonian along the RG ray $f_1 = - f_2 = f_3 = -f_4 = f_5 =  f > 0$ is mapped onto the chiral GN model (\ref{GN}).
Phase II is thus a C1S0 phase where all excitations are fully gapped except the uniform charge degrees of freedom.
The interacting Hamiltonian which governs the physical properties of this phase can be bosonized in close parallel to 
phase I and we find:
\begin{equation}
\begin{split}
&  {\cal H}_{\rm int}^{(II)} =  - \frac{f}{N} \left( \partial_x \Phi_{oL} \partial_x \Phi_{oR} 
- \partial_x {\vec \Phi}_{+L} \cdot  \partial_x {\vec \Phi}_{+R}  \right. \\
& \left. +  \partial_x {\vec \Phi}_{-L} \cdot \partial_x  {\vec \Phi}_{-R} \right)  \\
& + \frac{f}{2 \pi N a_0^2} \sum_{\alpha}  
\cos \left( \sqrt{\frac{8 \pi}{N}} \Theta_o + \sqrt{8 \pi} {\vec e}_{\alpha} \cdot {\vec \Theta}_{-}  \right) \\
& + \frac{f}{2 \pi N a_0^2} \sum_{\alpha \ne \beta}  \Gamma_{\alpha \beta} 
\cos \left( \sqrt{2 \pi} \left( {\vec e}_{\alpha} -  {\vec e}_{\beta} \right) \cdot {\vec \Phi}_{+}  \right) \\
& \times \left[ \cos \left( \sqrt{2 \pi} \left( {\vec e}_{\alpha} -  {\vec e}_{\beta} \right)  \cdot {\vec \Theta}_{-} \right) \right. \\
& \left.
- \cos \left( \sqrt{\frac{8 \pi}{N}} \Theta_o + \sqrt{2 \pi} \left( {\vec e}_{\alpha} +  {\vec e}_{\beta} \right)
\cdot {\vec \Theta}_{-} \right) \right],
\end{split}
\label{bosonizedHregionIIbis}
\end{equation}
and the pinning configurations for the bosons in phase II are now ($\Gamma_{\alpha \beta} = -1$) for instance:
$\langle \Theta_o \rangle = \sqrt{N \pi/8}, \langle {\vec \Phi}_{+}  \rangle = {\vec 0}$,  and
$\langle {\vec \Theta}_{-}  \rangle = {\vec 0}$.

Using the bosonization approach, the relative CDW operator (\ref{CDW-}) has the low-energy description:
 \begin{equation}
 \begin{split}
  {\cal O}_{{\rm CDW}_-}  & \simeq  \frac{1}{\pi a_0}  \sum_{\alpha} i \kappa_{1 \alpha} \kappa_{2 \alpha}  \cos \left( 2k_F x + \sqrt{\frac{2 \pi}{N}} \Phi_c \right. \\
 & \left. + \sqrt{2 \pi } {\vec e}_{\alpha}  \cdot {\vec \Phi}_{+}  \right) 
    \sin \left( \sqrt{\frac{2 \pi}{N}} \Theta_o + \sqrt{2 \pi } {\vec e}_{\alpha}  \cdot {\vec \Theta}_{-}  \right),
 \end{split}
\label{CDW-boson}
\end{equation}
so that by averaging over the massive excitations $ \Theta_o, {\vec \Phi}_{+}, {\vec \Theta}_{-}$, one gets:
 \begin{equation}
 {\cal O}{_{\rm CDW}}_-  \simeq  \frac{C}{\pi a_0}  \sum_{\alpha} i \kappa_{1 \alpha} \kappa_{2 \alpha}  \cos \left( 2k_F x + \sqrt{\frac{2 \pi}{N}} \Phi_c \right),
\label{CDW-bosonfin}
\end{equation}
where $C$ is a non-universal constant related to the different vacuum expectation values of the operators with 
the pinned bosons $\langle \Theta_o \rangle = \sqrt{N \pi/8}, \langle {\vec \Phi}_{+}  \rangle = {\vec 0}$, and
$\langle {\vec \Theta}_{-}  \rangle = {\vec 0}$. One obtains an order parameter with algebraic two-point correlation
function:
\begin{equation}
\langle  {\cal O}_{{\rm CDW}_{-}} (x)    {\cal O}_{{\rm CDW}_{-}} (0) \rangle \simeq  \frac{N^2 C^2}{2 \pi^2 a_0^2}  \frac{ \cos ( 2k_F x)}{x^{\frac{K_c}{N}}},
\label{CDW-correl}
\end{equation}
which has the same exponent as the OAF$_{\perp}$ order parameter (\ref{ODWycorrel}) of phase I. 

Similarly, the continuous description of the order parameter  ${\cal O}_{{\rm BDW}_{-}}$ can be bosonized in phase II 
and we obtain after averaging over the pinned bosonic fields:
\begin{equation}
\begin{split}
{\cal O}_{{\rm BDW}_-}   \simeq  \frac{C}{4\pi a_0} &  \sum_{\alpha} i \kappa_{1 \alpha} \kappa_{2 \alpha} 
\left[ e^{i 2k_F x + i \sqrt{\frac{2\pi}{N}} \Phi_c}  \left(e^{i k_{2F} a_0}   \right.  \right.\\
& \left. \left. + e^{i k_{1F} a_0}\right)  + {\text H.c.} \right],
\end{split}
\label{BDW-bosonfin}
\end{equation}
with equal-time correlation:
\begin{equation}
\langle {\cal O}_{{\rm BDW}_-} (x)   {\cal O}_{{\rm BDW}_-} (0) \rangle  \simeq  
 \frac{N^2 C^2}{2 \pi^2 a_0^2}\left(1-  \sin^2(\frac{\delta}{2}) \right)\frac{ \cos ( 2k_F x)}{x^{\frac{K_c}{N}}} .
 \label{BDW-correl}
\end{equation}
The CDW$_{-}$ correlation function (\ref{CDW-correl}) has a larger amplitude and the CDW$_{-}$  phase is stabilized  
for negative $U$ as it can be observed from  Fig. \ref{figinc}.

\section{Concluding remarks}
\label{sec:conclusion}

In the present paper, we have developed a low-energy description of weakly-coupled half-filled SU($N$) two-leg fermionic 
ladder. We have identified two distinct non-perturbative duality symmetries relating four competing orders.  The first one,
the density-current duality symmetry ${\cal D}_{\rm dc}$, is an exact symmetry of the lattice Hamiltonian that was previously introduced for the specific $N=2$ case in Refs. \onlinecite{Momoi-H-03,Momoi-H-05}. The latter duality symmetry naturally extends to arbitrary $N$ and maps  conventional phases, such as CDW$_{-}$ and BDW$_{-}$  ordered phases, onto unconventional T-breaking ordered phases characterized by spontaneous charge currents circulating in a staggered pattern either around the plaquettes or along diagonals of the ladder. In stark contrast, the second duality symmetry ${\cal D}$ emerges only in the low-energy limit. Under its action,  the two loop-current (respectively density) ordered phases are dual to each other. These orders can be continuously transformed into each other and vanish at a self-dual quantum 
critical point, where an emergent U(1) symmetry rotates the corresponding dual order parameters.

A one-loop RG approach of a generalized SU($N$) two-leg fermionic ladder describes the stabilization 
of the two loop-current and the CDW$_{-}$ and BDW$_{-}$  ordered phases as asymptotic RG rays characterized by
an enlarged SO($4N$) symmetry  in the far IR regime. The low-energy physics of these four competing orders are governed by the SO($4N$) GN model which is a massive integrable field theory.  As a consequence, all four phases correspond to fully gapped Mott-insulating C0S0 phases with a two-fold ground-state degeneracy.  We have found that
these phases are related to one another through  the hidden duality symmetries  ${\cal D}$  and  ${\cal D}_{\rm dc}$ of the SO($4N$) GN model. Interestingly, we have shown for $N>2$ that the four competing orders arise in the zero-temperature phase diagram of a half-filled SU($N$) two-leg Hubbard ladder with an additional SU($N$) Hund's coupling, i.e., an SU($N$) interchain spin-spin exchange interaction.
 
For an incommensurate filling close to half-filling,  we have investigated the effect of doping on these four phases 
by means of a one-loop RG approach. The  loop-current phases are characterized by algebraically decaying 
correlations with a $2k_F$ wave vector. This instability dominates other competing orders, including a subleading $4k_F$ CDW  that coexists with the loop-current phases.

An interesting perspective will be the numerical investigation of the lattice models considered here by means of large-scale SU($N$) DMRG calculations \cite{Nataf-25} in order to determine the extension of the loop-current ordered phases at half-filling and for generic fillings. In particular, the commensurate filling of one fermion per site is important as it is especially relevant for ultracold alkaline-earth and ytterbium atom experiments where three-body losses are suppressed for this filling.
In a future work, we plan to address the zero-temperature phase diagram of the ladder for such a filling within the weak-coupling regime.

\begin{acknowledgements}
The authors are very grateful to S. Capponi and K. Totsuka for very helpful discussions.
We would like  to thank CNRS for  financial support of the IRP project ``Exotic Quantum Matter in Multicomponent Systems (EXQMS)'' from  CNRS.  
\end{acknowledgements}

\appendix

\section{Abelian bosonization approach of the two-leg SU($N$) fermionic ladder}
\label{bosonisationappen}

We give here our conventions for the bosonization approach of the two-leg SU($N$) fermionic ladder.

From the massless Dirac fermions of the continuum limit (\ref{contlimitDirac}), we introduce left and right moving bosons 
$\varphi_{l \alpha}$ from the identification: \cite{Gogolin-N-T-book,James-K-L-R-T-18}
\begin{equation}
\begin{split}
 L_{l \alpha} &= \frac{\kappa_{l \alpha}}{\sqrt{2 \pi a_0}} \; e^{ - i \sqrt{4 \pi}\varphi_{l\alpha L}},  \\
 R_{l\alpha} &= \frac{\kappa_{l\alpha}}{\sqrt{2 \pi a_0}} \; e^{ i \sqrt{4 \pi}\varphi_{l \alpha R}}  ,
\end{split}
\label{bosoabeleq}
\end{equation}
where $l=1,2$ and $\alpha = 1, \ldots, N$ and  $[\varphi_{l \alpha R}, \varphi_{m \beta L} ] = i 
\delta_{l m} \delta_{\alpha\beta}/4$. In Eq. (\ref{bosoabeleq}),
$\kappa_{l\alpha}$ are Klein factors to insure  the anticommutation of fermions: $\{\kappa_{l \alpha}, \kappa_{m \beta} \} = 2   \delta_{l m } \delta_{\alpha\beta} $, and $\kappa^{\dagger}_{l \alpha} = \kappa_{l \alpha}$. 
Our normalization conventions of the chiral bosons are:
 \begin{eqnarray}
 \langle \varphi_{l \alpha L} (z) \varphi_{m \beta L} (0)\rangle &=& - \frac{\delta_{l m}\delta_{\alpha\beta}}{4 \pi} \ln z ,\nonumber \\
 \langle \varphi_{l \alpha R} (\bar z) \varphi_{m \beta R} (0)\rangle &=& - \frac{\delta_{l m}\delta_{\alpha\beta}}{4 \pi} \ln \bar z ,
 \label{bosonormalization}
\end{eqnarray}
with $z = v_{F} \tau + i x$.

The non-interacting massless Dirac Hamiltonian (\ref{HamcontDirac}) can be expressed in terms of these  bosonic
fields:
\begin{equation}
\begin{split}
&  {\cal H}_0= v_{F}  \sum_{l=1}^{2}\sum_{\alpha=1}^{N}    \left(:(\partial_x \varphi_{l\alpha L})^2: + 
  :(\partial_x \varphi_{l\alpha R})^2: \right) \\
 & =  \frac{v_{F}}{2} \sum_{l=1}^{2}\sum_{\alpha=1}^{N}    \left(:(\partial_x \varphi_{l\alpha})^2: + 
  :(\partial_x \vartheta_{l\alpha})^2: \right) ,
\end{split}
\label{HamDiracbos}  
\end{equation}  
where $\varphi_{l\alpha} = \varphi_{l\alpha L} + \varphi_{l\alpha R}$ is the total bosonic field and
$\vartheta_{l\alpha} =  \varphi_{l\alpha L} - \varphi_{l\alpha R}$ is its the dual field.

For each chain $l=1,2$, we introduce a charge bosonic field $\Phi_{lcL,R}$ and 
$N-1$ spin fields $\Phi_{lmL,R}, m=1, \ldots N-1$  which describes respectively charge and spin collective fluctuations.
In terms of the original chiral bosons, they are given by:\cite{Assaraf-A-C-L-99}
\begin{eqnarray}
&& \Phi_{lcR,L} = \frac{1}{\sqrt{N}} \sum_{\alpha =1}^{N} \varphi_{l\alpha R,L }, \label{SUNbasisapp} \\
&& \Phi_{lm R,L} =  \frac{1}{\sqrt{m(m+1)}} \left( \sum_{p=1}^{m} \varphi_{l pR,L} - m
\varphi_{lm+1 R,L} \right)  \nonumber .
\end{eqnarray}
The inverse transformation is:
\begin{eqnarray}
 \varphi_{l\alpha R,L }  &=& \frac{\Phi_{lcR,L}}{\sqrt{N}} + \sum_{m=1}^{N-1} e^{m}_{\alpha} \Phi_{lm R,L} 
 \nonumber \\
 &=&   \frac{\Phi_{lcR,L}}{\sqrt{N}} + {\vec  e}_{\alpha} \cdot {\vec  \Phi}_{lR,L}  ,
\label{invSUNapp}
\end{eqnarray}
where ${\vec  e}_{\alpha}$ ($\alpha=1, \ldots, N$)  are $N-1$-dimensional vectors  which satisfy:
\begin{subequations}
\begin{align}
\sum_{\alpha=1}^N \vec{e}_\alpha &= {\vec 0}, \\
\sum_{\alpha=1}^N [\vec{e}_\alpha]^m [\vec{e}_\alpha]^{m'} &= \delta_{mm'}, \\
 \vec{e}_\alpha {\cdot} \vec{e}_\beta
&= \delta_{\alpha \beta} -\frac{1}{N},
\end{align}
\label{weightSUN}
\end{subequations}
where $m=1, \ldots, N -1$ describes the components of the ${\vec  e}_{\alpha}$ vectors. An explicit choice is for instance:
\begin{align} \label{eq:SpecificWeight}
[\vec{e}_\alpha]^m = \begin{cases} \frac{1}{\sqrt{m(m+1)}} & (m \geq \alpha) \\ -\sqrt{\frac{m}{m+1}} & (m=\alpha-1) \\ 0 & (m<\alpha-1) \end{cases}. 
\end{align}
For instance, in the $N=3$ case we have:
\begin{align}
\begin{split}
{\vec e}_1 &= \begin{pmatrix} \frac{1}{\sqrt{2}}, & \frac{1}{\sqrt{6}} \end{pmatrix} ,\\
{\vec e}_2 &= \begin{pmatrix} -\frac{1}{\sqrt{2}}, & \frac{1}{\sqrt{6}} \end{pmatrix} , \\
{\vec e}_3 &= \begin{pmatrix} 0, & -\sqrt{\frac{2}{3}}  \end{pmatrix} .
\end{split}
\label{weightSU3}
\end{align}

With these definitions, the non-interacting Hamiltonian~(\ref{HamDiracbos}) reads as follows in the new basis: 
\begin{equation}
\begin{split}
&  {\cal H}_0=  \frac{v_{F} }{2} \sum_{l=1}^{2}   \left(:(\partial_x \Phi_{lc})^2: + 
  :(\partial_x \Theta_{lc})^2:  \right.\\
 & \left. + :(\partial_x {\vec \Phi_{l}})^2: + :(\partial_x {\vec \Theta_{l}})^2:
  \right) ,
\end{split}
\label{HamDiracbos2}  
\end{equation}  
where we have introduced the full bosonic field and its dual field for each degree of freedom.

It is useful to consider symmetric and antisymmetric combinations of the bosons to describe the collective excitations of the different degrees of freedom of the two-leg ladder: uniform charge, relative charge (or orbital) and spins. In this respect,  we introduce the following combination of chiral fields:
\begin{equation}
\begin{split}
& \Phi_{c L,R} = \frac{1}{\sqrt{2}} \left(\Phi_{1c L,R} + \Phi_{2 c L,R} \right), \\
& \Phi_{o L,R} = \frac{1}{\sqrt{2}} \left(\Phi_{1c L,R} - \Phi_{2 c L,R} \right) , \\
& {\vec \Phi}_{\pm L,R} = \frac{1}{\sqrt{2}} \left({\vec \Phi}_{1 L,R}  \pm {\vec \Phi}_{2 L,R} \right) .
\end{split}
\label{finalfields}
\end{equation}
We can then relate the initial chiral bosonic fields $\varphi_{l \alpha L,R}$ to the new ones (\ref{finalfields}) through:
\begin{equation}
\begin{split}
& \varphi_{l \alpha L,R} = \frac{1}{\sqrt{2N}}  \Phi_{c L,R} + \frac{1}{\sqrt{2N}} (-)^{l+1}  \Phi_{o L,R} \\
& +  \frac{1}{\sqrt{2}}  {\vec  e}_{\alpha} \cdot {\vec \Phi}_{+ L,R} 
+ \frac{1}{\sqrt{2}} (-)^{l+1}  {\vec  e}_{\alpha} \cdot {\vec \Phi}_{- L,R}  .
\end{split}
\label{finalident}
\end{equation}

The non-interacting Hamiltonian (\ref{HamDiracbos}) can be expressed in terms of these new bosonic fields: 
\begin{equation}
\begin{split}
  {\cal H}_0 &=  \frac{v_{F} }{2} \left(:(\partial_x \Phi_{c})^2: + 
  :(\partial_x \Theta_{c})^2:  \right) \\
 & +  \frac{v_{F} }{2} \left(:(\partial_x \Phi_{o})^2: + 
  :(\partial_x \Theta_{o})^2:  \right) \\
  & +  \frac{v_{F} }{2} \sum_{m = \pm} \left( :(\partial_x {\vec \Phi_{m}})^2:   + :(\partial_x {\vec \Theta_{m}})^2: \right) .
\end{split}
\label{HamDiracbos3}  
\end{equation}

\section{Duality symmetries and $\mathbb{Z}_{2}$-gradings of $\mathfrak{so}(4N)$ }
\label{dualityappen}

The duality symmetries found in this paper correspond to $\mathbb{Z}_{2}$-gradings of $\mathfrak{so}(4N)$ which are
 involutive automorphisms $\omega$ such that $\mathfrak{so}(4N) = \mathfrak{g}_{\parallel} \oplus  \mathfrak{g}_{\perp}$ with $\omega(X)=X$ (respectively $\omega(X)=-X$)  $\forall X\in \mathfrak{g}_{\parallel}$ (respectively 
$\mathfrak{g}_{\perp}$). These $\mathbb{Z}_{2}$-gradings of $\mathfrak{so}(4N)$ are known to be classified into two symmetric classes: the symmetric classes DIII  and BDI of the general classification of symmetric spaces. For DIII (respectively BDI), one has $\mathfrak{g}_{\parallel} = \mathfrak{u} (2N)$
(respectively  $\mathfrak{g}_{\parallel} = \mathfrak{so}(2N)  \oplus    \mathfrak{so}(2N)$) and 
$\mathfrak{g}_{\perp} = \mathfrak{so}(4N) /\mathfrak{u} (2N)$ (respectively 
$\mathfrak{g}_{\perp} = \mathfrak{so}(4N) / \mathfrak{so}(2N)  \oplus    \mathfrak{so}(2N)$).

In the low-energy approach, developped in this paper, we have found three duality symmetries ${\cal D}$, 
${\cal D}_{\rm dc}$ and ${\cal D}{\cal D}_{\rm dc}$. In the main text, we have seen that the duality 
${\cal D}$ is characterized by $\mathfrak{g}_{\parallel} = \mathfrak{u} (2N)$ and is  a $\mathbb{Z}_{2}$-grading
which belongs to the symmetric class DIII. In this Appendix, we determine the class of the 
$\mathbb{Z}_{2}$-gradings of $\mathfrak{so}(4N)$ corresponding to the remaining duality symmetries
${\cal D}_{\rm dc}$ and ${\cal D}{\cal D}_{\rm dc}$.

\subsection{The density-current duality ${\cal D}_{\rm dc}$ symmetry as a $\mathbb{Z}_{2}$-grading of $\mathfrak{so}(4N)$}

We first introduce $4N$ Majorana fermions from the $2N$ Dirac fermions of the continuum limit (\ref{contlimitDirac})
as in Eq. (\ref{majo}):
\begin{equation}
L_{l \alpha}  = ( \xi_{l \alpha L} + i \chi_{l \alpha L})/\sqrt{2}  .
\label{majoappen}
\end{equation}
The density-current duality ${\cal D}_{\rm dc}$ symmetry on the Dirac fermions (\ref{densitycurrduality}) leads to the following transformation on the Majorana fermions:
\begin{equation}
\begin{split}
\xi_{l \alpha L}   & \xrightarrow{{\cal D}_{\rm dc}}   (-)^{l} \chi_{l \alpha L}, \\
\chi_{l \alpha L}   & \xrightarrow{{\cal D}_{\rm dc}}   (-)^{l+1} \xi_{l \alpha L}  ,
\end{split}
\label{dcdualmajo}
\end{equation}
the right-moving Majorana fermions being invariant under the transformation.  
We introduce two new $2N$ left-moving Majorana fermions:
\begin{align}
\eta_L =  \begin{pmatrix}
         \xi_{1 \alpha L} \\
         \; \;  -  \xi_{2 \alpha L} 
    \end{pmatrix}   ,     \\
\rho_L =  \begin{pmatrix}
         \chi_{1 \alpha L} \\
         \; \;   \chi_{2 \alpha L} 
    \end{pmatrix}        ,
\label{2Nmajodualdc}
    \end{align}
so that under the duality symmetry ${\cal D}_{\rm dc}$, we have:
\begin{equation}
\begin{split}
\eta_{L}   & \xrightarrow{{\cal D}_{\rm dc}}   -  \rho_L ,  \\
\rho_{L}   & \xrightarrow{{\cal D}_{\rm dc}}    \eta_L   .
\end{split}
\label{dcdualmajo2N}
\end{equation}
The SO($4N$)$_1$ left current can be defined in terms of a $4N$-component Majorana fermion $\gamma_L$ 
which is built from the two $2N$ Majorana fields $\eta_{L}$ and $\rho_{L}$: $\gamma_L = (\eta_{L}, \rho_{L} )^{T}$. The SO($4N$)$_1$  current
reads then as follows
\begin{equation}
I^{A}_L = - i  \gamma_{a L} \gamma_{b L} ,
\label{dcdualmajo}
\end{equation}
with $ A= (a,b)$ and  $1 \le a < b \le 4N$. These currents satisfy the SO($4N$)$_1$ current algebra \cite{DiFrancesco-M-S-book}:
\begin{equation}
I^{A}_L (z)  I^{B}_L (0) \sim  \frac{ \delta^{AB}}{ 4 \pi^2 z^2} + i  f^{ABC} \frac{ I^{C}_L (0)}{2 \pi z},
\label{so4N_1algebra}
\end{equation}
where $f^{ABC}$ are the structure constants of  $\mathfrak{so}(4N)$. 

The duality symmetry ${\cal D}_{\rm dc}$ is a $\mathbb{Z}_{2}$-grading of $\mathfrak{so}(4N)$ with
the decomposition $\mathfrak{so}(4N) = \mathfrak{g}_{\parallel} \oplus  \mathfrak{g}_{\perp}$. The SO($4N$)$_1$ currents
(\ref{dcdualmajo}) can then be decomposed into even and odd sectors $I^{A}_L = (I^{B}_{\parallel L}, I^{C}_{\perp L})$ under ${\cal D}_{\rm dc}$:
\begin{equation}
\begin{split}
I^{B}_{\parallel L}  & \xrightarrow{{\cal D}_{\rm dc}}   I^{B}_{\parallel L}, \;  B = 1, \ldots, {\rm dim}   
\mathfrak{g}_{\parallel} , \\
I^{C}_{\perp L}  & \xrightarrow{{\cal D}_{\rm dc}}   -  I^{C}_{\perp L}, \;  C = 1, \ldots, {\rm dim}   
\mathfrak{g}_{\perp}  ,
\end{split}
\label{Z2gradingSO(4N)_1}
\end{equation}
which enables us to identify the symmetry class of the density-current duality symmetry. In this respect, we combine the two $\eta_L$ and $\rho_L$ Majorana fermions into a $2N$-component Dirac fermion as in Eq. (\ref{majoappen}):
\begin{equation}
\Psi_{m L}  = ( \eta_{m L} + i \rho_{m L})/\sqrt{2}  ,
\label{majobisappen}
\end{equation}
with $m=1, \ldots, 2N$. The next step of the approach is to define an U($2N$)$_1$ current from these Dirac fermions as in Eq. (\ref{SU(2N)_1curr}):
\begin{equation}
\begin{split}
\mathcal{J}_{L}^{A} & =  \Psi_{m L}^{\dagger} {\cal T}^A_{ m n} \Psi_{n L} ,  \\
\mathcal{J}_{L}^{0} & =  :\Psi_{m L}^{\dagger} \Psi_{m L}: ,
\end{split}
\label{U(2N)_1curr}
\end{equation}
with $A=1, \ldots, 4N^2-1$, $m,n = 1, \ldots, 2N$, and ${\cal T}^A$  are the SU($2N$) generators in the fundamental representation of the SU(2$N$) group with the normalization: ${\rm Tr}( {\cal T}^A {\cal T}^B) = \delta^{AB}/2$. 
The latter generators can be classified  into three categories: 
\begin{itemize}
\item {\rm Antisymmetric, i.e., SO($2N$) part:}
\begin{equation}
\left({\cal T}_{\rm{A}}^{(a,b)}\right)_{m n}
=-\frac{i}{2}(\delta_{a m}\delta_{b n}
-\delta_{a n}\delta_{b m})
\qquad (1\leq a < b \leq 2 N)
\label{eqn:SU2N-gen-1}
\end{equation}
\item {\rm Symmetric part:}
\begin{equation}
\left({\cal T}_{\rm{S}}^{(a,b)}\right)_{m n}
=\frac{1}{2}(\delta_{a m}\delta_{b n}
+ \delta_{a n}\delta_{b m})
\qquad (1\leq a < b \leq 2N)
\label{eqn:SU2N-gen-2}
\end{equation}
\item {\rm Cartan generators (Diagonal):}
\begin{equation}
 \left({\cal T}_{\rm{D}}^{a}\right)_{m n}
=\frac{1}{\sqrt{2 a(a+1)}}
\left(
\sum_{k=1}^{a}\delta_{m k}\delta_{n k}
-a \, \delta_{m,a+1}\delta_{n,a+1}
\right),
\label{eqn:SU2N-gen-3}
\end{equation}
\end{itemize}
with $a=1,\ldots,2N-1$.

The U($2N$)$_1$  currents of Eq. (\ref{U(2N)_1curr}) can then be decomposed as:
\begin{itemize}
\item {\rm Antisymmetric}
\begin{equation}
I^{(a,b)}_{{\rm A} \parallel L} 
= - \frac{i}{2} \left(  \eta_{a L}   \eta_{b L}  + \rho_{a L} \rho_{b L} \right)
\qquad (1\leq a < b \leq 2 N)
\label{antisymmcurr}
\end{equation}
\item {\rm Symmetric}
\begin{equation}
I^{(a,b)}_{{\rm S} \parallel L} 
= \frac{i}{2} \left(  \eta_{a L}   \rho_{b L}  + \eta_{b L} \rho_{a L} \right)
\qquad (1\leq a < b \leq 2N)
\label{symmcurr}
\end{equation}
\item {\rm U(1)}$^{2N}$ {\rm commuting part:}  ($a = 1, \ldots, 2N-1$)
\begin{equation}
\begin{split}
I^{a}_{ \parallel L} 
& = \frac{i}{\sqrt{2 a(a+1)}}
\left(
\sum_{m=1}^{a } \eta_{m L}   \rho_{m L} 
-a \, \eta_{a+1 L}   \rho_{a+1 L} 
\right) , \\
I^{0}_{ \parallel L} 
& =  i \sum_{m=1}^{2N}  \eta_{m L}   \rho_{m L}  .
\end{split}
\label{commutingcurr}
\end{equation}
\end{itemize}
Using the transformation (\ref{dcdualmajo2N}) on the Majorana fermions, it is easy to check that the U($2N$)$_1$  currents (\ref{antisymmcurr}, \ref{symmcurr}, \ref{commutingcurr}) are invariant under the density-current duality symmetry ${\cal D}_{\rm dc}$. These currents form the subset of SO($4N$)$_1$ currents (\ref{dcdualmajo}) which are
even under ${\cal D}_{\rm dc}$. The remaining $2N(2N -1)$ SO($4N$)$_1$  currents are given by:
\begin{equation}
\begin{split}
I^{(a,b)}_{{\rm A} \perp L} & = - \frac{i}{2} \left(  \eta_{a L}   \eta_{b L}  - \rho_{a L} \rho_{b L} \right) , \\
I^{(a,b)}_{{\rm S} \perp L} & = - \frac{i}{2} \left( \eta_{a L}   \rho_{b L}  - \eta_{b L} \rho_{a L}   \right),
\end{split}
\label{perpso4N1current}
\end{equation}
with $1\leq a < b \leq 2 N$. These currents are odd under ${\cal D}_{\rm dc}$. The density-current duality 
symmetry ${\cal D}_{\rm dc}$ gives thus a $\mathbb{Z}_{2}$-grading of the current algebra SO($4N$)$_1$ 
(\ref{Z2gradingSO(4N)_1}) with $\mathfrak{g}_{\parallel} = \mathfrak{u}(2N)$. It belongs thus to the 
DIII symmetry class of $\mathbb{Z}_{2}$-gradings of $\mathfrak{so}(4N)$.

\subsection{The duality symmetry ${\cal D}{\cal D}_{\rm dc}$ as a $\mathbb{Z}_{2}$-grading of $\mathfrak{so}(4N)$}

We now consider the last duality symmetry which is the combination of ${\cal D}$ and ${\cal D}_{\rm dc}$. 
Its effect of the Majorana fermions introduced in Eq. (\ref{majoappen}) is:
\begin{equation}
\begin{split}
\xi_{l \alpha L}   & \xrightarrow{{\cal D} {\cal D}_{\rm dc}}   (-)^{l} \xi_{l \alpha L} , \\
\chi_{l \alpha L}   & \xrightarrow{{\cal D} {\cal D}_{\rm dc}}   (-)^{l} \chi_{l \alpha L}  ,
\end{split}
\label{ddcdualmajo}
\end{equation}
the right-moving Majorana fermions being invariant.  As before, we introduce two $2N$ left-moving Majorana fermions:
\begin{align}
\eta_{lL} =  \begin{pmatrix}
         \xi_{l \alpha L} \\
         \; \;  \chi_{l \alpha L} 
    \end{pmatrix}      , 
\label{2Nmajodualddc}
\end{align}
with $l=1,2$ and under the duality symmetry ${\cal D}  {\cal D}_{\rm dc}$, we have:
\begin{equation}
\begin{split}
\eta_{1 L}   & \xrightarrow{{\cal D}  {\cal D}_{\rm dc}}   -  \eta_{1 L} ,  \\
\eta_{2 L}   & \xrightarrow{{\cal D}  {\cal D}_{\rm dc}}   \eta_{2 L}   .
\end{split}
\label{ddcdualmajo2N}
\end{equation}
The SO($4N$)$_1$ currents (\ref{dcdualmajo}) can be separated into two parts with different parities with
respect to the duality symmetry ${\cal D}  {\cal D}_{\rm dc}$. The even sector is formed by:
\begin{equation}
\begin{split}
I^{(a,b)}_{1 \parallel L} & = - i  \eta_{1 a L}   \eta_{1b L} ,  \\
I^{(a,b)}_{2 \parallel L} & = - i  \eta_{2 a L}   \eta_{2 b L} ,
\end{split}
\label{parallelso4N1current}
\end{equation}
with  $1\leq a < b \leq 2 N$ and they describe the free-field representation of
two commuting SO($2N$)$_1$ currents. Using Eq. (\ref{ddcdualmajo2N}), one observes that these currents  are even under the duality symmetry ${\cal D}{\cal D}_{\rm dc}$. 
The remaining currents are given by:
\begin{equation}
I^{(a,b)}_{\perp L}  =   - i  \eta_{1 a L}   \eta_{2 b L}, \qquad (1\leq a \leq b \leq 2N) ,
\label{perprcurrDDdc}
\end{equation}
which are odd under the duality symmetry ${\cal D}{\cal D}_{\rm dc}$. The latter transformation defines
a $\mathbb{Z}_{2}$-grading of the current algebra SO($4N$)$_1$ 
(\ref{Z2gradingSO(4N)_1}) with $\mathfrak{g}_{\parallel} = \mathfrak{so}(2N)  \oplus    \mathfrak{so}(2N)$. The duality symmetry ${\cal D}{\cal D}_{\rm dc}$belongs thus to the BDI symmetry class of $\mathbb{Z}_{2}$-gradings of $\mathfrak{so}(4N)$.


%

\end{document}